\documentclass[12pt]{article}

\usepackage[left=1in, top=1.2in, right=1in, bottom=1.5in]{geometry}

\usepackage{tabularx}
\usepackage{graphicx}
\usepackage{booktabs}
\usepackage{todonotes}
\usepackage{setspace}

\usepackage[flushleft]{threeparttable}

\usepackage{xr} 

\usepackage{amsmath,amsfonts,amssymb}
\usepackage{amsthm}

\usepackage[natbib,style=authoryear,
uniquename=false,
maxcitenames=2,maxbibnames=99,
uniquelist=false,
giveninits=true,
doi=false,isbn=false,
url=false]{biblatex}
\usepackage{bibentry}
\renewbibmacro{in:}{}
\renewbibmacro*{volume+number+eid}{\printfield{volume}\setunit*{\addnbspace}\printfield{number}\setunit{\addcomma\space}\printfield{eid}}
\DeclareFieldFormat[article]{number}{\mkbibparens{#1}}
\DeclareNameAlias{author}{family-given}

\newcommand{\comment}[1]{}  

\usepackage{eurosym}
\usepackage{tikz}
\usepackage{pdflscape}
\usepackage{footnote}

\usepackage{bbm}
\usepackage{enumitem}   

\usepackage{subcaption}

\makeatletter
\g@addto@macro{\table}{\centering \small}
\makeatother

\makeatletter
\g@addto@macro{\figure}{\centering \small}
\makeatother

\DeclareCaptionFont{mysize}{\fontsize{10}{12}\selectfont}
\newtheorem{definition}{Definition}
\newtheorem{assumption}{Assumptions}
\newtheorem{lemma}{Lemma}

\newcommand{\dE}{\mathbb{E}}
\newcommand{\dP}{\mathbb{P}}

\newcommand{\ind}[1]{\ensuremath{\mathbbm{1}}(#1)}

\DeclareMathOperator*{\argmin}{arg\,min}

\newcommand{\disas}{\sim}
\DeclareMathOperator*{\argmax}{arg\,max}

\newcommand{\red}[1]{#1}

\newcommand{\out}[1]{}

\providecommand{\keywords}[1]
{
	\small	
	\textbf{\textit{Keywords---}} #1
}

\newcommand{\bM}{[M_q]}
\newcommand{\bE}{[E_q]}
\newcommand{\bC}{[C_q]}

\newcommand{\defeq}{\mathrel{\mathop:}=}

\newcommand{\du}{u_w}

\newcommand\restr[2]{{
		\left.\kern-\nulldelimiterspace 
		#1 
		\vphantom{\big|} 
		\right|_{#2} 
}}

\newcommand{\av}{a}
\newcommand{\bv}{b}

\newcommand{\pstar}{p^*(x)}
\newcommand{\bi}{B}

\def\home{.}

\title{Eliciting ambiguity with mixing bets}
\date{This version: \today}
\author{Patrick Schmidt\footnote{\textit{Address for			correspondance:} Patrick Schmidt, Binzmühlenstrasse 14, 8050 Zurich, Switzerland. E-mail: pschmidte@gmail.com. I would like to thank Aurelien Baillon, Arup Daripa, Markus Eyting, Luca Henkel, Toru Kitagawa, Michael Kosfeld, Charles Manski, Matthias Schündeln, and Peter Wakker for insightful discussions. My work has been partially funded by the Klaus Tschira Foundation.}\\ University of Zurich}

\begin{document}
	\clearpage	\maketitle
	\thispagestyle{empty}
		
	\begin{abstract}
Preferences for mixing can reveal ambiguity perception and attitude on a single event. The validity of the approach is discussed for multiple preference classes including maxmin, maxmax, variational, and smooth second-order preferences. An experimental implementation suggests that participants perceive almost as much ambiguity for the stock index and actions of other participants as for the Ellsberg urn, indicating the importance of ambiguity in real-world decision-making.
\end{abstract}


\clearpage

\setcounter{page}{1}

Subjective expected utility (SEU) is commonly the basis of economic modeling \citep{savage1972foundations}. However, uncertainty often cannot be represented by a precise probability measure. Initiated by \citet{ellsberg1961risk}, various experiments showed that ambiguity matters in decision-making.
While \textit{artificially} generated ambiguity in experiments is well studied, there is little evidence of ambiguity for \textit{natural} uncertainty \citep{wakker2018measuring}, which complicates real-world applications of ambiguity-sensitive models. In this paper, I propose a simple mechanism - \emph{mixing bets} - that reveals subjective ambiguity for a natural event. I show that mixing bets are informative under a wide range of models and provide experimental evidence on their applicability.

\subsection{Separation of Ambiguity Perception: The Belief Interval}\label{sec:intro.bi}

To understand the empirical content of decision models, it is crucial to separate perception and attitude \citep{manski2004measuring}. This is especially important for decisions under ambiguity, where typically no rational benchmark exists. The large number and flexibility of ambiguity-sensitive models make this separation challenging \citep[see the discussion in][]{epstein2010paradox,klibanoff2012smooth}. Following \citet{klibanoff2014perceived}, ambiguity perception can be represented by the set of probabilities that possibly govern the occurrence of an event. I call the interval that contains all such probabilities the \emph{belief interval}. Preferences are said to exhibit \emph{ambiguous beliefs} if the belief interval is not a single point. In the following, I define the belief interval for specific preferences and show how to identify said interval with mixing bets. 

Consider an act $l$ that depends on an uncertain event $E$. The following representations allow defining a belief interval. The classical subjective expected utility (SEU) by \citet{savage1972foundations} can be represented with a single probability $p$ in the unit interval and a utility function $u$ by
$$\dE_p[u(l)].$$
Ambiguity-sensitive models cannot be described with a single probability. Maxmin expected utility (maxmin) by \citet{gilboa1989maxmin} can be represented with a belief interval $\bi=[a,b]$ by
$$\min_{p \in \bi} \dE_p[u(l)].$$

The more general variational preferences by \citet{maccheroni2006ambiguity} can be represented with a positive cost function $c$ by
$$\min_{p \in \bi} \dE_p[u(l)]+c(p),$$
\red{where the belief interval $B$ is the subset of the unit interval, where the cost function $c$ is finite.}

Smooth second-order preferences by \citet{klibanoff2005smooth} can be represented with a probability measure $\dP$ on the unit interval and a second-order utility function $\phi$ by
$$\dE_{p \disas \dP}[ \phi( \dE_p[u(l)])].$$
For second-order preferences, the belief interval $\bi$ is the support of the probability measure $\dP$.

\subsection{Mixing Bets}\label{sec:intro.mixing}

To reveal the belief interval and ambiguity attitude, I propose the following mechanism. The agent is endowed with lottery tickets, each representing a fixed probability of winning a prize (e.g., a monetary reward). The agent has to bet each ticket on the event or its complement. The two events have different betting odds. After one of the two events realizes, the agent receives the tickets bet on the realized event times the odds of the realized event. Let $q\in(0,1)$ denote the standardized odds. Then, the procedure can be described as follows:
\begin{enumerate}
	\item The agent chooses the ratio $x$ of lottery tickets that she bets on the event $E$ (and the remainder $1-x$ on its complement $E^c$).
	\item If the event $E$ realizes, the agent receives the prize with probability $xq$. If the complement $E^c$ realizes, the agent receives the prize with probability $(1-x)(1-q)$.
\end{enumerate}
This task is called a \emph{mixing bet}. The agent's response will depend on $q$. The lottery tickets guarantee robustness to the unknown utility function \citep{smith1961consistency} if the randomization device is perceived as an independent lottery. In addition, the procedure can be repeated with different odds $q$, where one instance is randomly selected for payout to prevent hedging across the repeated betting tasks. \red{The validity of this approach will be discussed later.}

\subsection{Illustration: A Simple Mixing Bet}\label{sec:intro_easy}

\begin{figure}
	\includegraphics[width=\textwidth]{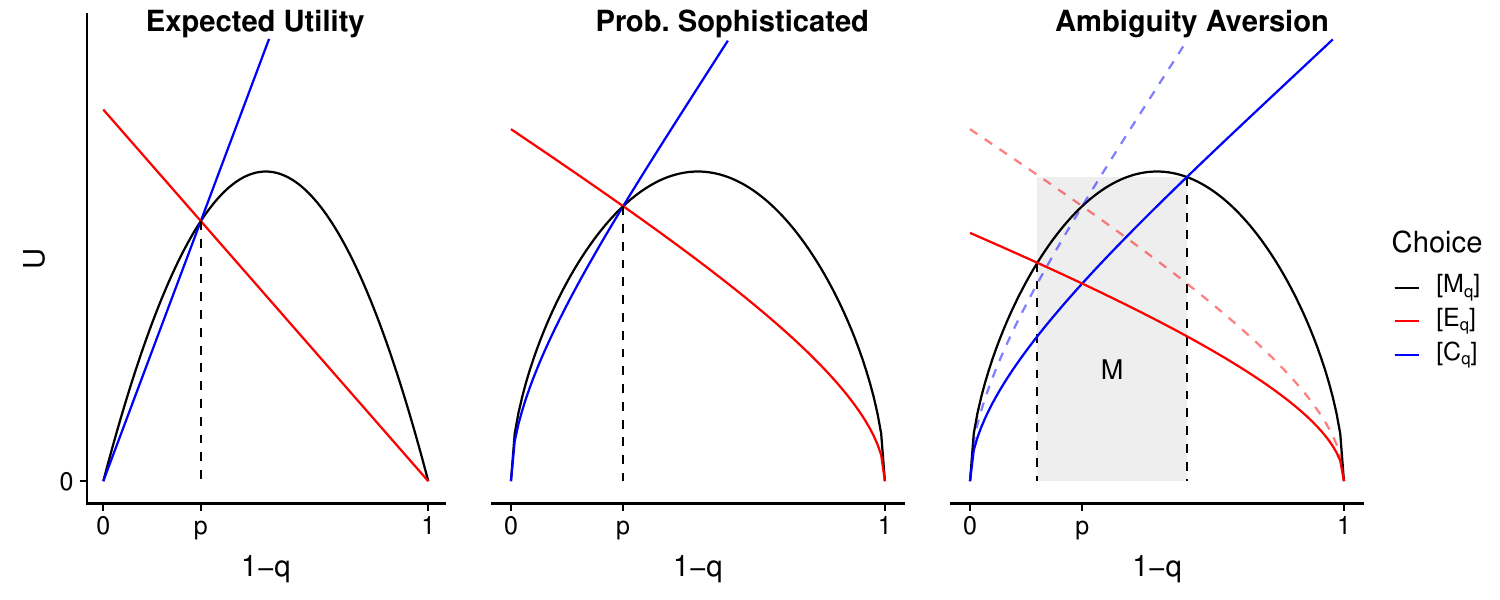}
	\caption{The value of the three choices $\bE,\bC$, and $\bM$ depending on $1-q$. The left plot illustrates the utility functional under SEU, the middle plot under probabilistically sophisticated preferences \citep{machina1992more}, and the right plot under ambiguity-averse preferences. In this example, the subjective probability is $p=0.3$, and the probabilistically sophisticated values are based on a probability weighting function $w(p)= \exp(-(-\ln(p))^{3/4})$ \citep[compare ][]{prelec1998probability}. Under ambiguity aversion, the ambiguous choices $\bE$ and $\bC$ are less attractive, and $\bM$ is strictly preferred for any $1-q$ in the interval $M$.
	}
	\label{fig:values_easy}
\end{figure}

First, consider an illustration of how mixing can reveal ambiguity aversion. Take the three focal choices of a mixing bet
\begin{itemize}
	\item[$\bE$] a lottery that pays with probability $q$ if the event $E$ realizes (``betting all tickets the event''),
	\item[$\bC$] a lottery that pays with probability $1-q$ if the event $E$ does not realize (``betting all tickets on the complement''), and
	\item[$\bM$] a lottery that pays with probability $q(1-q)$ (``mixing'').
\end{itemize}
Note that $\bE$, $\bC$, and $\bM$ arise for the choices $x=1$, $x=0$, and $x=1-q$ respectively, where $\bM$ can be interpreted as a probabilistic mixture of option $\bE$ and option $\bC$ that does not depend on the potentially ambiguous event $E$.
Figure \ref{fig:values_easy} illustrates the value associated with each choice for SEU preferences, probabilistically sophisticated preferences \citep{machina1992more}, and ambiguity-averse preferences. All three examples assign the ambiguity-neutral probability $p$ to the event $E$. Under all preferences, option $\bC$ becomes more attractive and option $\bE$ less attractive with increasing $1-q$. The value $1-q$ at which the decision-maker switches between the choice $\bE$ and $\bC$ can be used to elicit the subjective probability $p$.

Under SEU, the value of $\bE$ and $\bC$ is linear in $q$. Further, the decision-maker is indifferent between the mixture $\bM$ and its elements $\bE$ and $\bC$ if those have equal value. \red{There exists no $q$ such that the mixing choice $\bM$ is strictly preferred to $\bE$ and $\bC$. The same holds for probabilistically sophisticated preferences, where values are monotonically transformed and preferences over the three choices remain unaffected. }

Under ambiguity aversion, the value of the mixing choice $\bM$ remains unchanged, the ambiguous choices $\bE$ and $\bC$, however, are less attractive. \red{Thus, the mixing choice $\bM$ is strictly preferred for some interval $M$ of values $1-q$.}
In particular, the set $M$ contains the probability $p$ associated with ambiguity-neutral preferences. Further, the set $M$ is larger for more ambiguity-averse preferences.

\subsection{Theoretical Results}\label{sec:intro.theoretical_results}

\red{I establish the following results for the ambiguity-averse preferences introduced in Section \ref{sec:intro.bi}: If $1-q$ is above the belief interval, the agent bets all tickets on the complement. Reversed, if $1-q$ is below, the agent bets all tickets on the event. 
Mixing (betting tickets on the event \emph{and} the complement) is a sufficient condition for $1-q$ being in the belief interval. Beliefs are ambiguous (i.e., they do not reduce to a single probability level) if and only if the agent mixes for at least two different values of $q$.} 

The interval of values of $1-q$ that induces mixing, called the \textit{mixing interval}, can be used as a proxy for ambiguity perception. The mixing interval lies within the belief interval, such that the belief interval can be bounded from within. For maxmin preferences, the bounds are sharp. Under second-order and variational preferences with sufficiently strong ambiguity aversion, the bounds become sharper for large utility differences between the prizes.

Mixing choices also reveal information about ambiguity attitude. \red{If the agent hedges fully, i.e. $x=1-q$, the payoff does not depend on the ambiguous event. The mixing intensity, i.e., how close choices are to this full hedging, reveals aspects of ambiguity attitude with maxmin (and Choquet) preferences inducing the most intense mixing and variational and second-order preferences inducing less intense mixing.}

Further, I propose a generalization of mixing bets with two separate choices that reveals the belief interval not only for ambiguity-averse preferences but also for ambiguity-seeking preferences.

\red{The assumptions in the theoretical part abstract from hedging across the random payment scheme, noise, probability weighting on lotteries, and aversion to compound objects. I discuss the potential impact of those phenomena and propose a stochastic choice model based on generalized preferences that can adjust measurements.}

\subsection{Experimental Evidence}

In a laboratory experiment, the mechanism is applied to events generated by an Ellsberg urn, a stock index, and another participant's behavior in a prisoner's dilemma game. \red{I propose three indices to quantify different aspects of ambiguity-sensitive models. The \emph{midpoint of the mixing interval} measures the ambiguity-neutral probability. The \emph{length of the mixing interval} reflects ambiguity perception. Finally, the \emph{mixing intensity} provides additional information on ambiguity attitude.}

The experiment provides some evidence for the validity of \red{the} indices. Ambiguity perception is highest for the ambiguous and lowest for the risky category in the Ellsberg urn. After observing additional draws from the urn, the measured ambiguity decreases. Interestingly, ambiguity perception for the \emph{natural} events, generated by the stock index and another participant's behavior, is almost as large as for the Ellsberg urn. This suggests that ambiguity can be important in real-world decision-making.

The experimental evidence suggests that the mixing intensity is an individual-specific property that provides additional information next to the mixing interval. \red{Only a small share of participants behaves consistently with the extreme mixing predicted by maxmin preferences. Instead, participants often preferred more moderate mixing, which is consistent with variational or second-order preferences.}

Experience with uncertainty is negatively correlated with ambiguity perception for the stock index and the risky draw from the urn, but not with ambiguity perception for the ambiguous draw or the other participants' behavior. Cooperation in the prisoner's dilemma is correlated with the midpoint of the mixing interval (indicating reciprocity) but not with risk aversion, ambiguity aversion, or ambiguity perception. 

\red{
	Additionally, I estimate a stochastic choice model accounting for hedging, noise, probability weighting, and aversion to compound objects. The model is estimated in a Bayesian framework and provides evidence for all phenomena. The main take-away of the simpler analysis, however, remains unchanged: Mixing bets can distinguish between risk and ambiguity, and ambiguity perception for the natural events is almost as high as for the Ellsberg urn.}

\subsection{Related Literature}

The most closely related literature uses matching probabilities \citep{dimmock2015ambiguity,baillon2015testing} to elicit ambiguity.
\citet{wakker2018measuring} propose \emph{belief hedges} to construct indices of ambiguity attitude and perception based on the matching probabilities of three mutually exclusive events and their pairwise unions. \citet{baillon2018balanced} show that their indices are insightful under a wide range of ambiguity-sensitive models.\footnote{It is noteworthy, however, that theoretical results for the validity of belief hedges under second-order preferences are limited to events constructed as a mixture between the ambiguous event and a risky option with a vanishing ambiguous part \citep[Lemma~33,][]{baillon2018balanced}. Further, theoretical validation for variational preferences remains absent, except for the subset of multiplier preferences \citep[Equ.~15,][]{baillon2018balanced}.}

	Belief hedges have been applied productively in the literature. \citet{li2018trust} apply the method in an adapted trust game where the second player has an additional third option. \citet{anantanasuwong2019ambiguity} elicit ambiguity perception about different assets from a sample of investors by dividing the return values into three intervals.
	\citet{henkel2022experimental} experimentally investigates the ambiguity perception index after eliciting matching probabilities based on three intervals of temperature. Note that all applications relied on three events, where a single event (and its complement) would have been sufficient.

	While the \emph{index of ambiguity aversion} \citep[Def.~6,][]{baillon2018balanced} is applicable for a single event, the \emph{index of ambiguity-generated-insensitivity} \citep[Def.~10,][]{baillon2018balanced} requires at least three events. For a single event, mixing bets provide richer information by revealing the mixing interval and the mixing intensity simultaneously. As the standard Ellsberg urn draw is just a single event, the \textit{index of ambiguity-generated-insensitivity} cannot be validated in this standard setting. I provide empirical validation for the mixing bets indices by revisiting the Ellsberg urn. In many applications, only a single event is of interest (e.g., becoming jobless, bank default, regime change, nuclear war) and the construction of three mutually exclusive events is an unnecessary complication. 

Other related elicitation work obtained identification results at the expense of generality across decision models or the simplicity of the mechanism. \citet{bose2017eliciting} extend the mechanism introduced by \citet{karni2009mechanism} to $\alpha$-maxmin preferences. In another paper, \citet{bose2017secondorder} introduce a mechanism that identifies the distribution of beliefs for second-order preferences. \red{Recently, \citet{bleichrodt2023testing} showed how to elicit a two-parameter weighting function that can be interpreted as ambiguity perception and ambiguity aversion for Hurwicz expected utility preferences \citep{gul2015hurwicz}. The preferences allow for probability weighting for risk and do not impose universal ambiguity aversion or seeking, accounting for commonly observed empirical deviations from SEU \citep[compare the arguments in][]{bleichrodt2023testing}. The method is applicable for real-valued variables based on the exchangeability method \citep{abdellaoui2011rich} or matching probabilities \citep{dimmock2015ambiguity}.}
Another recent contribution, proposes to elicit matching \textit{intervals of probabilities} \citep{hill2021eliciting}, which can be interpreted as the belief interval and provides a solution to the identifiability problem in $\alpha$-maxmin preferences \citep{hill2023beyond}. Earlier related work includes \citet{dow1992uncertainty}, who identify an interval of prices for which an investor neither buys nor sells under Choquet expected utility \citep{schmeidler1989subjective}.

Applied work often considers the relation between the object of interest and ambiguity attitude measured on artificial events like the Ellsberg urn. See for example \citet{muthukrishnan2009ambiguity} for consumer and \citet{bianchi2019ambiguity} for portfolio choice. \citet{vives2018tolerance} find that prosocial behavior in a prisoner's dilemma is correlated with ambiguity tolerance measured on an artificial event. I find no correlation between cooperation in the prisoner's dilemma and ambiguity measured on the actually relevant event (the other participant's behavior). Instead, the ambiguity-neutral probability, measured by the midpoint of the mixing interval, is the most relevant predictor suggesting reciprocity to be more important than ambiguity.

\red{In Section \ref{sec:theory} the mixing behavior under different preferences is derived.} Sections \ref{sec:theory}.1 to \ref{sec:theory}.3 cover maxmin, variational, and second-order preferences respectively. Section \ref{sec:seeking} introduces a generalization, \emph{separated mixing bets}, which reveal ambiguity-seeking and ambiguity-averse preferences simultaneously. 
Section \ref{sec:experiment} provides an experimental implementation and empirical evidence. 
Section \ref{sec:robust} discusses robustness and Section \ref{sec:discussion} concludes. Proofs are provided in the appendix. 
Other ambiguity-averse preferences, e.g. biseparable preferences \citep{ghirardato2001risk} that include $\alpha$-maxmin \citep{marinacci2002probabilistic,ghirardato2004differentiating} and Choquet expected utility \citep{schmeidler1989subjective}, do not allow for a similar separation of a belief interval from ambiguity attitude.
A supplementary document discusses mixing bets under those preferences and under general ambiguity-averse preferences \citep{cerreia2011uncertainty}.

\section{Optimal Mixing under Ambiguity Aversion} \label{sec:theory}

Consider the task of eliciting beliefs about an event $E$ from an agent with unknown preferences. The state space is given by $S=\{E,E^c\}$, where any state $s\in S$ describes the realization of the event $E$. Following the \citet{anscombe1963definition} framework, the agent's preferences $\succeq$ are defined on acts 
that assign to each state a lottery over money rewards. 
Formally, a mixing bet with choice $x$ can be associated with such an act. For any $p\in[0,1]$, let $[p]^w_0$ denote the lottery that pays out a prize $w$ with probability $p$ and $0$ with probability $1-p$. 
The act resulting from a mixing bet with prize $w$, odds $q$ and response $x$ can be described as
$$l_{x,q,w}(s)
= [\ind{s=E}xq+\ind{s=E^C}(1-x)(1-q)]_0^w.$$

The mixing interval $M$ describes all odds $q$ for which the agent prefers mixing, $x\in (0,1)$, to betting fully on the event, $x=1$, or the complement, $x=0$.

\begin{definition}[mixing interval $M$]
	Let $x^\ast(q) \subset [0,1]$ define the best response for a mixing bet with value $q$ such that
	$$l_{x^\ast(q),q,w} \succeq l_{x,q,w} \text{ for all $x \in [0,1]$.}$$
	The mixing interval $M$ is defined as the smallest closed interval that contains
	$$\{q \in [0,1] \;|\; x^\ast(1-q) \neq \{1\} \text{ and } x^\ast(1-q) \neq \{0\}  \}.$$
\end{definition}

The mixing intensity measures how close mixing choices are to full hedging.

\begin{definition}[mixing intensity]
	For any value $q$ in the mixing interval $M$ and mixing choice $x$, the \textit{mixing intensity} $i$ is given by $i(x,q)\defeq \min(x/(1-q),(1-x)/q)$.
	For a set of mixing choices, the \textit{mixing intensity} is defined as the average of the mixing intensity $i(x(q),q)$.
\end{definition}
Note that the mixing intensity is picewise linear in $x$ and obtains its minimum $0$ for $x=0$ and $x=1$ and its highest value $1$ at $x=1-q$.

The belief interval definitions in Section \ref{sec:intro.bi} capture perceived ambiguity with the range of relevant probabilities following \citet{klibanoff2014perceived}.\footnote{\citet{klibanoff2014perceived} provide a behavioral definition of relevant probabilities in smooth models that coincides with the belief interval for maxmin and second-order preferences.} Heuristically, the belief interval $\bi$ denotes the relevant probabilities $p$ that the agent considers when making decisions related to the uncertain event $E$. Sections \ref{sec:theory.seu} to \ref{sec:theory_secondorder} provide details on the uniqueness of the belief interval. For simplicity, I consider only belief intervals instead of arbitrary sets.

In the following, the best responses to a mixing bets are considered for different preferences.

\subsection{Subjective Expected Utility}\label{sec:theory.seu}

First, consider the best response to the betting mechanism for an agent with SEU preferences.

\begin{assumption}[SEU]\label{ass:seu}
	The agent has SEU preferences with a belief $p \in [0,1]$ about the event $E$. In particular, the preferences can be represented by
	$$U(l)= \dE_{s \disas p}[u(l(s))]$$
	for some strictly increasing utility function $u$.
\end{assumption}  

The expected utility is linear in $x$ such that the agent prefers to bet everything on one event, where the switching point depends on the subjective probability $p$.

\begin{figure}
	\includegraphics[width=.3\textwidth]{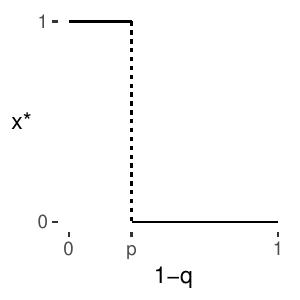}
	\caption{Optimal response for SEU preferences with belief $p=0.3$.}
	\label{fig:seu}
\end{figure}

\begin{lemma}[SEU]\label{th:betseu}
	The optimal choice for SEU preferences in Assumptions \ref{ass:seu} is
	$$x^\ast(q)=\begin{cases}
		1  &\text{if  } \; p > 1-q\\
		[0,1]  &\text{if  } \; p = 1-q\\
		0 &\text{if  } \; p < 1-q.\\
	\end{cases}$$
\end{lemma} 

The proof of Lemma \ref{th:betseu} is given in the appendix. 
The optimal choices for SEU preferences are illustrated in Figure \ref{fig:seu}. Mixing is optimal if and only if $1-q$ equals the subjective probability $p$. If the elicitor observes $x^\ast(q)=1$, it follows that $p>1-q$. For $x^\ast(q)=0$, it follows that $p<1-q$. Thus, choices for different values of $q$ reveal an interval that contains the belief $p$.

\subsection{Maxmin Preferences}\label{sec:maxmin}

This section establishes the optimal mixing for maxmin preferences with belief interval $\bi$.
\begin{assumption}[maxmin]\label{ass:maxmin}
	The agent holds maxmin preferences with belief interval $\bi=[a,b]$ about the event $E$. In particular, the preferences can be represented by
	$$U(l)= \min_{p \in \bi} \dE_{s \disas p}[u(l(s))]$$
	for some strictly increasing utility function $u$.
\end{assumption}  

The set of measures $\bi$ is unique \citep[][Theorem~1]{gilboa1989maxmin} and the belief interval is well-defined. As a special case, maxmin preferences contain SEU preferences if the beliefs are unambiguous with $\bi=\{p\}$.

\begin{lemma}[maxmin]\label{th:betmaxmin}
	The optimal choice for maxmin preferences as in Assumptions \ref{ass:maxmin} is
	\[
	x^\ast(q)= \left\{\begin{alignedat}{3}
		&1     &&\text{if } \hphantom{a <{}} 1-q < a\\
		&1-q  \quad &&\text{if } {a} < 1-q < b \\
		&0   &&\text{if } b < 1-q
	\end{alignedat}\right. 
	\]
\end{lemma} 

Lemma \ref{th:betmaxmin} follows from the more general statement for variational preferences in Lemma \ref{th:betsvariationallimited}. See Lemma \ref{th:betsalphamaxmin} in the supplementary document for $\alpha$-maxmin preferences.

\begin{figure}
	\includegraphics[width=.3\textwidth]{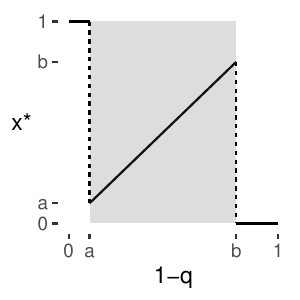}
	\caption{Optimal response for maxmin preferences with ambiguous belief interval  $\bi=[a, b]$. The shaded area marks the belief interval, which is identified by the mixing behavior.}
	\label{fig:maxmin}
\end{figure}

Figure \ref{fig:maxmin} illustrates the best response under maxmin preferences. \red{The mixing interval $M$ equals the belief interval $B$. The mixing intensity is $1$.} Note that the extreme ambiguity attitude inherent in those preferences results in a linearly increasing response on the belief interval $[a,b]$. Interpreting betting behavior for maxmin preferences is straightforward. If everything is betted on the complement $E^c$, the belief interval $\bi$ is below $1-q$. If everything is betted on $E$, the belief interval is above $1-q$. Finally, if mixing is observed, the belief interval contains $1-q$.

\subsection{Variational Preferences}\label{sec:theory_variational}

This section establishes the mixing behavior under variational preferences \citep{maccheroni2006ambiguity}, which generalize multiplier preferences \citep{hansen2007beliefs,hansen2007recursive}. We assume variational preferences with belief interval $\bi$.

\begin{assumption}[variational preferences]\label{ass:limitedVP}
	The agent has variational preferences. In particular, the preferences over acts $l(E)$ can be represented by
	$$U(l)=  \min_{p \in \bi}  \dE_{s \disas p}[u(l(s))] + c(p)$$
	for some strictly increasing utility function $u$ and some grounded, strictly convex and twice continuously differentiable cost function $c:\bi\to [0,\infty)$.
\end{assumption}  
The cost function $c$ is grounded such that there exists $p \in B$ with $c(p)=0$. All variational preferences are ambiguity-averse, with maxmin preferences as a special case. Assumptions \ref{ass:limitedVP} cover variational preferences as defined in \citet{maccheroni2006ambiguity} if they are twice continuously differentiable and strictly convex on $\{p \in [0,1] \;|\; c(p)< \infty\}$. To see this, extend the cost function $c$ to the unit interval with a slight abuse of notation and define $c(p)=\infty$ for $p \notin \bi$. Given a utility function $u$, the minimal $c$ is unique \citep[Theorem~3,][]{maccheroni2006ambiguity} and the belief interval $\bi$ is defined by the closure of $\{p \in [0,1] : c(p)<\infty \}$. Rescaled utility functions represent the same preferences when paired with a rescaled cost function $c$ \citep[Corollary~5,][]{maccheroni2006ambiguity}, such that the belief interval is unique.

\begin{figure}
	\includegraphics[width=.6\textwidth]{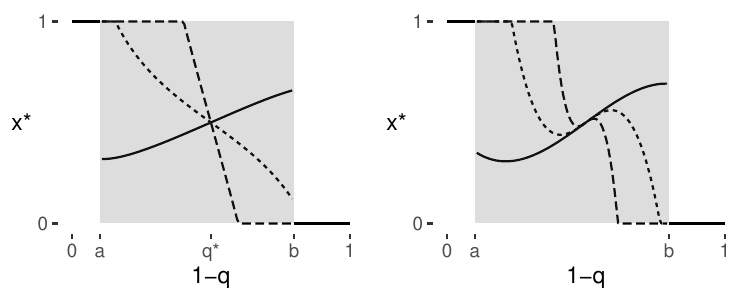}
	\caption{Optimal response for variational preferences. The optimal choice is continuous in $q$. Six examples are shown, where the belief interval is $\bi=[0.1,0.8]$ throughout. The left plot depicts multiplier preferences with $c(p)=\theta R(p||0.5)$ and $\theta=0.1, 0.5, 1.5$, where $R$ is the relative entropy function. The right plot depicts variational preferences with $c(p)=\theta |p-0.5|^4$ and $\theta=1,10,100$.}
	\label{fig:variational}
\end{figure}

\begin{lemma}[variational preferences]\label{th:betsvariationallimited}
	If the agent follows variational preferences as in Assumptions \ref{ass:limitedVP} with belief interval $\bi=[a,b]$, the optimal choice for a mixing bet with prize $w$ is
	\[
	x^\ast(q)= \left\{\begin{alignedat}{3}
		&1     &&\text{if } \hphantom{a <{}} 1-q < a\\
		&f_w(q)  \quad &&\text{if } {a} \le 1-q \le b \\
		&0   &&\text{if } b < 1-q
	\end{alignedat}\right. 
	\]
	for a continuous function $f_w:B\to[0,1]$ and it holds that 
	\begin{enumerate}[label=(\roman*)]
		\item $f_w(q)=1-q$ if and only if $c(1-q)=0$,
		\item the mixing interval is $M_{\du}=\{q\in B | c'(1-q)/\du<1-q<1+c'(1-q)/\du\}$,
		\item on $M_{\du}$ the mixing function is $f_w(q)=1-q-\tfrac{c'(1-q)}{\du}$, and
		\item if $c'$ is bounded and $\du$ is unbounded, there exists a prize $w$ such that the mixing interval identifies the belief interval, $M_{\du}=\bi.$
	\end{enumerate}
\end{lemma} 

The rich behavior of mixing bets under variational preferences can be summarized as follows. Mixing is never optimal outside of the belief interval. The mixing intensity is $1$ for $c(1-q)=0$, which allows to reveal the interval $\{p\in B| c(p)=0\}$. The remainder of the mixing interval exhibits less severe mixing with $x^\ast(q)\neq 1-q$. If the utility difference $\du$ is sufficiently large, the mixing interval recovers the belief interval.

Note that the mixing intensity reveals aspects of the ambiguity attitude as captured in the function $c$. Given a fixed utility function $u$, variational preferences become more ambiguity averse with smaller cost functions $c$ \citep[compare Proposition~8,][]{maccheroni2006ambiguity}. Consider two agents with identical belief intervals and utility functions, where $c_2 = \lambda c_1$ for some $\lambda>1$ such that $c_2$ expresses less ambiguity aversion. The more ambiguity-averse agent mixes more intensely, i.e. $x^\ast(q)$ is closer to $1-q$, which allows us to order agents by ambiguity attitude.

\subsection{Smooth Second-Order Preferences}\label{sec:theory_secondorder}

This section considers smooth second-order preferences \citep{klibanoff2005smooth}.

\begin{assumption}\label{ass:smooth}
	The agent holds beliefs $\dP$ in form of a distribution over $[0,1]$ with support $\bi=[a,b]$ with $0\le a \le b \le 1$ about the event $E$ and has ambiguity-averse smooth second-order preferences. In particular, the preferences over acts $l(E)$  can be represented by
	$$U(l)=  \dE_{p \disas \dP}[ \phi( \dE_{s \disas p}[u(l(s))])]$$
	for some strictly increasing utility function $u$ and some strictly increasing, concave, and twice continuously differentiable second-order utility function $\phi$.
\end{assumption}  

The second-order probabilities $\dP$ are almost surely unique and the belief interval is unique across representations. See \citet{klibanoff2014perceived} for a theoretical discussion on capturing the perception of ambiguity under second-order preferences and beyond.

\begin{lemma}\label{th:betssmooth}
	The optimal choice for a mixing bet with prize $w$ of an ambiguity-averse agent with second-order preferences as in Assumptions \ref{ass:smooth} is 
	\[
	x^\ast(q)= \left\{\begin{alignedat}{3}
		&1     &&\text{if } \hphantom{a <{}} 1-q \le a\\
		&g_w(q)  \quad &&\text{if } {a} < 1-q < b \\
		&0   &&\text{if } b \le 1-q,
	\end{alignedat}\right. 
	\]
	for some continuous function $g_w:B\to[0,1]$ with $g_w(1-q)<1$ if $1-q>\dE_{p \disas \dP}[p]$ and $g_w(1-q)>0$ if $1-q<\dE_{p \disas \dP}[p]$.\\
	In particular, it holds that $\dE_{p \disas \dP}[p] \in M_{\du}$. Further, if the coefficient of ambiguity aversion $\alpha(z)=-\tfrac{\phi''(z)}{\phi'(z)}$ is bounded away from zero, it holds that for every element $p\in (a,b)$ there exists a sufficiently large utility difference $\du$ such that $p\in M_{\du}$.
\end{lemma} 

\begin{figure}
	\includegraphics[width=.3\textwidth]{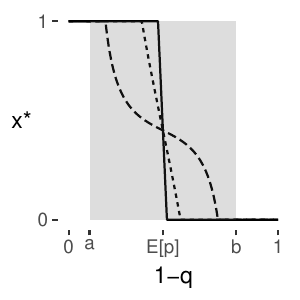}
	\caption{Optimal response for second-order ambiguity-averse preferences. The optimal choice is continuous in $q$ and lies in the shaded rectangle. Three examples are shown, where  $u(0)=0$, $u(w)=1$, the second-order distribution $\dP=U[0.1,0.8]$ and the second-order utility function is $\phi(z)=-e^{-\theta z}$ with $\theta=1,4,16$ respectively.}
	\label{fig:smoothjoint}
\end{figure}

The continuity of $g_w$ implies that the agent is mixing on an interval of positive length. For sufficiently strong ambiguity aversion second-order preferences are essentially identical to maxmin preferences \citep[][Proposition 3]{klibanoff2005smooth} and the belief interval $B=[a,b]$ can be identified with a high degree of accuracy. Lemma \ref{th:betssmooth} shows that the same effect can be induced by increasing the utility difference $\du$ if one is willing to assume strictly positive ambiguity aversion.

In Figure \ref{fig:smoothjoint} three examples with different constant absolute ambiguity aversion are shown. Bounds on the belief interval are conservative for moderate rates of ambiguity aversion and low utility difference $\du$ between prizes. \red{The exact determination of the belief interval with moderate utility differences $\du$ is complicated by the interaction between perception (represented by $\dP$) and attitude (represented by $\phi$ and $u$).}

While the interaction of higher moments of $\dP$ and higher derivatives of $\phi$ remain elusive, the main characteristics, namely the ambiguity-neutral probability $\dE_{p \disas \dP}[p]$ and some proxy for ambiguity perception, are available. Importantly, a typical assumption would be that attitude in form of $\phi$ and $u$ remains constant for an individual irrespective of learning or the target event $E$. Thus, the difference in mixing elicited from the same person for two events is informative about the difference in ambiguity perception.

\section{Ambiguity-Seeking Preferences}\label{sec:seeking}

The \textit{mixing bets} considered until here cannot identify ambiguity perception for ambiguity-seeking preferences. While ambiguity aversion is more prominent empirically than ambiguity seeking, it is not universal \citep{li2018rich}. This section introduces a generalization, \textit{separated mixing bets}, that allows distinguishing between ambiguity-seeking and ambiguity-averse preferences and identifying the belief interval \red{in both cases}.

\subsection{Illustration: Separated Mixing Bets}

\begin{figure}
	\includegraphics[width=\textwidth]{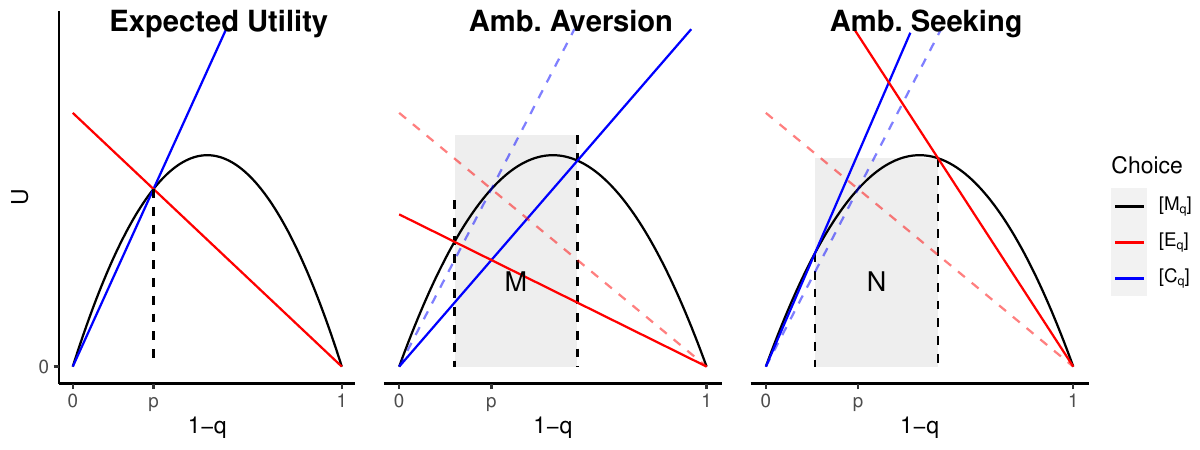}
	\caption{The value of the three choices $\bE,\bC$, and $\bM$ depending on $1-q$. The left plot illustrates the utility functional under SEU, the middle plot under ambiguity-averse, and the right plot under ambiguity-seeking preferences. Under ambiguity-seeking preferences, the ambiguous choices $\bE$ and $\bC$ are more attractive such that they are both preferred to the mixing choice $\bM$ for any $1-q$ in the interval $N$.
	}
	\label{fig:values_easyS}
\end{figure}

To build intuition, consider an extension of the illustration from Section \ref{sec:intro_easy} to ambiguity-seeking preferences as depicted in Figure \ref{fig:values_easyS}. Again, we analyze the value of the three choices $\bE$, $\bC$, and $\bM$ (betting on the event, the complement, and mixing) under different preferences.

First, we observe that mixing is never preferred under ambiguity-seeking preferences. The choice $\bM$ is dominated by $\bE$ (for low $1-q$), $\bC$ (for high $1-q$), or both (for intermediate $1-q$). The behavior for standard mixing bets would be indistinguishable from SEU.

Instead, the difference between SEU and ambiguity-seeking preferences arises if the choice between $\bM$ and its components is observed separately. An ambiguity-seeking agent prefers the ambiguous bets $\bE$ and $\bC$ over the unambiguous $\bM$ for some values of $1-q$. We refer to this interval as the \emph{non-mixing interval} $N$. 

\subsection{Separated Mixing Bets}

Separated mixing bets are a generalization of the mixing bets considered before. The agent chooses twice, where in each choice the options are limited to allocations of lottery tickets that favor one of the two outcomes. For a mixing bet with value $q$, the feasible allocations are restricted to $[1-q,1]$ (favoring the event) and $[0,1-q]$ (favoring the complement).

In this scenario, the mixing interval $M$ describes all values of $q$ for which the agent is mixing in both separated mixing bets. Additionally, the \textit{non-mixing interval} $N$ describes all values of $q$ for which the agent is not mixing in either of the two separated mixing bets. Note that the mixing interval recovers the information from the simple mixing bets considered before. The non-mixing interval provides additional information for ambiguity-seeking preferences.

\subsection{Maxmax Preferences}

This section establishes the optimal mixing in separated mixing bets for maxmax preferences with belief interval $\bi$. \red{(The according results for ambiguity-seeking smooth second-order preferences can be found in the Supplementary Section \ref{sec:smoothseeking}.) }Maxmax preferences arise as the most ambiguity seeking special case of $\alpha$-maxmin expected utility preferences \citep{marinacci2002probabilistic,ghirardato2004differentiating}.

\begin{assumption}[maxmax]\label{ass:maxmax}
	The agent holds maxmax preferences with belief interval $\bi=[a,b]$ about the event $E$. In particular, the preferences can be represented by
	$$U(l)= \max_{p \in \bi} \dE_{s \disas p}[u(l(s))]$$
	for some strictly increasing utility function $u$.
\end{assumption}  

\begin{figure}
	\includegraphics[width=.5\textwidth]{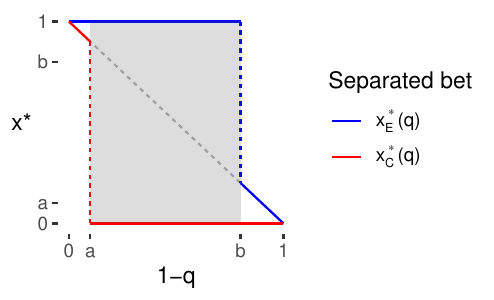}
	\caption{Optimal response for maxmax preferences with ambiguous belief interval  $\bi=[a,b]$. The shaded area marks the belief interval, which is identified by the mixing behavior.}
	\label{fig:maxmax}
\end{figure}

\begin{lemma}[maxmax]\label{th:betmaxmax}
	The optimal responses for the separated mixing bets for maxmax preferences as in Assumptions \ref{ass:maxmax} are
	\[
	x_E^\ast(q)= \left\{\begin{alignedat}{2}
		&1     &&\text{if }  1-q < b\\
		&1-q  \quad &&\text{if } 1-q > b \\
	\end{alignedat}\right.
	\quad \text{ and } \quad 
	x_C^\ast(q)= \left\{\begin{alignedat}{2}
		&0     &&\text{if } a < 1-q \\
		&1-q  \quad &&\text{if } a > 1-q \\
	\end{alignedat}\right.  
	\]
\end{lemma}

The results of Lemma \ref{th:betmaxmax} are illustrated in Figure \ref{fig:maxmax}. The proof follows from the arguments in the general statement in Lemma \ref{th:betsalphamaxmin} in the supplementary document for $\alpha$-maxmin preferences.

Interpreting betting behavior for maxmax preferences is straightforward. If the decision-maker is not choosing to mix for either of the two mixing bets, then $1-q$ is in the belief interval and the preferences are ambiguity seeking. If the decision-maker chooses to mix in both separated mixing bets, then $1-q$ is in the belief interval and the preferences are ambiguity averse. If the decision-maker chooses the pure event in one and mixes for the other bet, then $1-q$ is beyond the belief interval and no statements about the ambiguity attitude can be made.

\section{Experiment}\label{sec:experiment}

I provide empirical evidence that mixing bets are applicable, measure ambiguity, and can generate insightful evidence. The focus is on mixing bets, omitting separated mixing bets from Section \ref{sec:seeking}.

In the experiment, participants chose mixing bets on draws from an Ellsberg urn, on the stock market, and on another participant's behavior in a prisoner's dilemma. The comparison of mixing bet choices between the events can shed light on the validity of the mechanism and provide evidence for the relevance of ambiguity for natural sources of uncertainty.

\subsection{Setup}

For the experiment 88 student subjects were recruited at the Frankfurt Laboratory for Experimental Economic Research. The experiment was implemented with OTree.\footnote{Data and code for replication is available online \citep{schmidt2024replication}.} The average age was 25 and $49\%$ of participants were female. Most participants studied economics or business (38\%), mathematics or natural sciences (14\%), social science (14\%), or law (14\%).

Participants received 5 Euros for participation. First, they played a standard prisoner's dilemma with potential payoffs of 1 Euro (both defect), 2 Euros (both cooperate) and 0/3 Euros (cooperate/defect). Next, risk aversion was measured with a choice between 2 Euros, 5 Euros with a $50\%$ chance, and 10 Euros with a $30\%$ chance. The probabilities were generated by drawing a number from 1 to 100 from a box. The same mechanism was used subsequently as the lottery generating device for the mixing bets. Participants could win a prize of 10 Euros with the mixing bets. Before the mixing bets elicitation, participants received instructions, examples, and test questions for the payoff mechanism. At the end of the experiment, demographic data were collected, the realization of the events was shown and an envelope was opened that contained the mixing bet number that would determine the payout. Then, the number of obtained lottery tickets was shown on the screen and each participant had to draw a number from a box to determine the final payoff.

In the main part of the experiment, I revisit the Ellsberg urn with an urn that contains 90 balls, where the color composition (30 blue, 60 red) was known, but the number of dotted balls (0 - 60 with dots) was unknown. Further, the current value of the German Stock Index was written down at the beginning of the experiment (before the German Stock Exchange opened) and at the end of the experiment (about 30 minutes after trading started). 
For the mixing bets, participants considered five different events: Drawing a blue ball (\textit{risk}), drawing a dotted ball (\textit{ambiguity}), the stock market rising (\textit{stock}), the assigned player in the prisoner's dilemma choosing to defect (\textit{social}). The order of elicitation was randomized for each participant. Finally, all participants saw 10 draws from the Ellsberg urn and repeated the dotted ball elicitation (\textit{updated}). For each domain, the $q$ values $\{0.1,0.2,\dots, 0.9\}$ were applied, where the order was randomized, and the values above (below) were skipped if a higher (lower) $q$ values elicited the answer $x=0$ ($x=1$). If the procedure finished with four or fewer values of $q$ considered, an additional choice for another value of $q$ was elicited to check the consistency. Throughout, odds were scaled such that betting all tickets on one event implied certainty. 

I consider two ways of framing the elicitation with mixing bets. In the \textit{discrete elicitation}, participants encounter a series of pairwise choices representing the three extreme choices $x \in \{ 0, 1-q, 1\}$ from Section~\ref{sec:intro_easy}. A single decision in the \emph{discrete elicitation} is relatively simple as illustrated by the following example.\\

What do you prefer? 10 Euros if $\dots$
\begin{itemize}
	\item $\dots$ a red ball will be drawn and additionally you draw a number between 1 and 25.
	\item $\dots$ a blue ball will be drawn.
\end{itemize}

In the \textit{continuous elicitation} participants could choose $x$ with a slider, where for each slider position the payoff for the two events was shown on the screen (e.g., for the middle position $x=0.5$ with value $q=0.5$ the screen shows ``If the stock index rises, you will receive 10 Euros if you draw a number between 1 and 50.'' and ``If the stock index declines, you will receive 10 Euros if you draw a number between 1 and 50.''.). Feasible allocations were reduced to $x \in \{0, 0.1, \dots, 0.9,1\}$. The continuous elicitation followed after the discrete elicitation and to avoid excessive waiting times, it was skipped if a participant fell behind too much for the updated, stock, and social domain, but not for the risk and ambiguity domain. About 50\% of participants were subject to the continuous elicitation for all domains.

\subsection{Consistency of Responses}

First, we evaluate if mixing bets were conceptually understood by the participants. After the elicitation and before seeing the results, 78\% of participants affirmed that they ``completely understood the payout mechanism''. This subjective assessment was not correlated with gender, area of study, or experience in statistics. Another indicator is the consistency of answers. In total, $8\%$ of answers exhibit non-mixing between mixing values. Another $2\%$ bet on the event while betting on the complement for lower values of $q$. The remaining $90\%$ of answers are consistent with the choice patterns derived in Section \ref{sec:theory}.\footnote{Note that extremer values of $q$ were omitted after betting on the event/complement. Thus, not all such inconsistencies would have been revealed.} Comparing choices elicited in the continuous and discrete elicitation for the same domain and value of $q$, $4\%$ of choice pairs mix in the discrete, but prefer non-mixing in the continuous elicitation, and $12\%$ prefer non-mixing in the discrete elicitation, but prefer a choice indicating higher likelihood of the other event in the continuous elicitation. The remaining $84\%$ of paired choices are consistent with the patterns derived in Section \ref{sec:theory}. 

\subsection{Mixing Behavior}

Multiple mixing implies ambiguity aversion (Section \ref{sec:theory}). 82\% of participants show such behavior in at least one of the three domains ambiguity, social, and stock. Within-subject heterogeneity is prominent with only 32\% of participants exhibiting multiple mixing for all three domains.

For comparison, \citet{dominiak2011attitudes} find that only 16\% subjects value a mixed option (based on a coin flip) strictly higher than its Ellsberg-type ambiguous components. Considering matching probabilities, \citet{wakker2018measuring} report that less than half of a sample of Dutch university students showed signs of ambiguity aversion for the stock index and \citet{li2018trust} observe that about half of their participants were ambiguity averse regarding another participant's behavior in a trust game.

Ambiguity-seeking preferences imply non-mixing for all mixing bets\footnote{Ambiguity neutral preferences imply mixing for exactly one value of $q$. As only a finite number of values is elicited, non-mixing for all values is also consistent with ambiguity neutral preferences.}. Only $3\%$ of participants never preferred mixing, providing strong evidence against universal ambiguity seeking. However, 48\% of the participants did not mix for at least one of the three domains such that a moderate prevalence of source-specific ambiguity seeking cannot be ruled out. As \emph{separated mixing bets} were not elicited, the experimental setting cannot distinguish between ambiguity-seeking and ambiguity-neutral preferences.

\subsection{Indices of Ambiguity Perception and Aversion}\label{sec:indices}

To summarize the elicited mixing bets, I use three indices: the \emph{midpoint of the mixing interval}, the \emph{length of the mixing interval}, and the \emph{mixing intensity}. The indices can be derived directly from mixing choices and can be interpreted as ambiguity-neutral probability, ambiguity perception, and ambiguity aversion. 

The first index is the \emph{midpoint of the mixing interval}, or if no mixing occurred, the midpoint of the switching interval. It can be interpreted as the ambiguity-neutral probability. For maxmin and variational preferences, it approximates the midpoint of the belief interval and for second-order preferences the mean of the second-order distribution. 

The second index is the \emph{length of the mixing interval}, i.e. the difference between the highest and lowest $q$ that induced a mixing response. For maxmin preferences, it denotes the size of the belief interval (Lemma \ref{th:betmaxmin}). For variational preferences, it does so if the prize is sufficiently attractive (Lemma \ref{th:betsvariationallimited}). For second-order preferences, the separation of attitude and perception is more challenging and the length of the mixing interval identifies the size of the belief interval only for high stakes (Lemma \ref{th:betssmooth}). As we can only observe choices for a finite number of values of $q$, the measured length of the mixing interval is weakly smaller than its true length. The sharpness of the bounds can be controlled with the choice of mixing bets. With the values $q \in \{0.1,0.2,\dots, 0.9\}$, it follows that the difference is smaller than $0.2$.\footnote{Note that a similar issue arises for belief hedges \citep{wakker2018measuring}, where the underlying matching probabilities are identified only up to the interval of probabilities, where participants switch from the uncertain event to the risky option.} A mixing interval with positive length is evidence for ambiguity perception with ambiguity aversion.

The third index is the \emph{mixing intensity}, where extreme mixing is $x_q=1-q$ and the least mixing is $x_q=0.1$ or $x_q=0.9$. The mixing intensity is measured with the continuous elicitation as the discrete elicitation allowed only extreme mixing. For maxmin preferences, the intensity is $1$ (Lemma \ref{th:betmaxmin}). For variational preferences, the mixing intensity allows ranking cost functions by ambiguity aversion (Lemma \ref{th:betsvariationallimited}). Note, however, that outcome-dependent preferences require the utility difference to be adequate. Low utility differences induce no mixing and high utility differences induce strong mixing.

\begin{figure}
	\includegraphics[width=\textwidth]{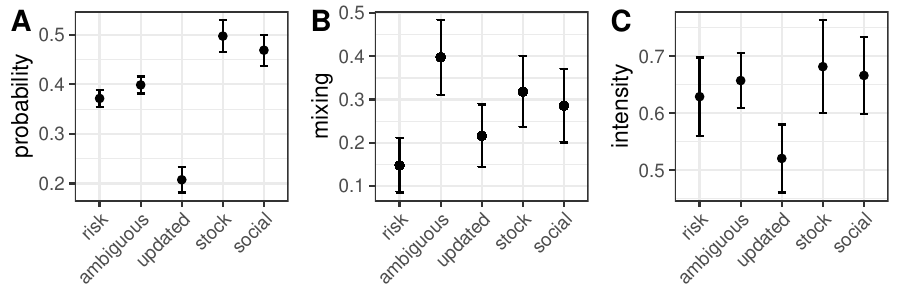}
	\caption{\textbf{Proxies for ambiguity and probability perception.} The plot depicts the average and the $90\%$-confidence intervals for the population mean of the indices. Panel A shows the the ambiguity-neutral probability, panel B the occurrence of a mixing interval with positive length, and panel C the mixing intensity. Panel A and B depend on the discrete and panel C on the continuous elicitation.}
	\label{fig:exp_averages}
\end{figure}

Figure \ref{fig:exp_averages} shows the average of the three proxies by domain. In panel A, the ambiguity-neutral probabilities are consistent with reality. The average for the risky urn is close to the true value $\tfrac{1}{3}$ and for the ambiguous urn only slightly higher. After the 10 additional draws, the average belief for the ambiguous category is updated in the direction of the true value $\tfrac{1}{10}$. The stock market is expected to rise with an average probability of $50\%$. In the cooperation game, participants expect their partner to defect with an average probability of $45\%$. The observed average was close with $46\%$ of participants defecting.

Panel B considers the likelihood of a mixing interval with positive length (evidence for ambiguity perception). Mixing is least prevalent for the risky urn (15\%) and most prevalent for the ambiguous urn (40\%) (but decreasing sharply after 10 draws were shown). The two natural events, \textit{stock} and \textit{social}, exhibit a likelihood of mixing of $32\%$ and $29\%$. That is lower than the ambiguous draw ($p$-value of $t$-tests: $0.27$ and $0.13$), but higher than the risky draw ($p$-value of $t$-tests: $0.01$ and $0.03$). \red{Section \ref{sec:robust} discusses several reasons for the risky urn draw to indicate non-zero ambiguity.}

The main goal of the experiment is to reveal ambiguity beyond the Ellsberg urn. The two examples considered here indicate that natural events, where uncertainty is generated by mechanisms beyond the experimental control, are indeed perceived as ambiguous by a considerable ratio of participants. Note that the German stock index should be familiar to the participant pool, which suggests even higher ambiguity perception for less commonly known stocks \citep{abdellaoui2011rich}.

In panel C, the mixing intensity is similar across sources, consistent with the idea that mixing intensity reveals a personal ambiguity attitude instead of ambiguity perception. The exception is the \emph{updated} domain, where the beliefs are closer to zero and the limited allocations of  $x \in \{0, 0.1, \dots, 0.9,1\}$ could distort measurements. \red{This finding is consistent with \citet{bleichrodt2023testing}, who find that ambiguity aversion measured based on the exchangeability method is constant across sources of uncertainty \citep[see also][]{li2018rich}.}

\begin{table}
	\centering
	\footnotesize
	\renewcommand\baselinestretch{1}\selectfont
	\begin{tabular}{@{\extracolsep{5pt}}lcccccc} 
\\[-1.8ex]\hline 
\hline \\[-1.8ex] 
 & \multicolumn{6}{c}{\textit{Dependent variable:}} \\ 
\cline{2-7} 
\\[-1.8ex] & mixing  & risk & \multicolumn{4}{c}{multiple mixing} \\
 & intensity & aversion & risk & ambiguity & stock & social \\ 
\\[-1.8ex] & (1) & (2) & (3) & (4) & (5) & (6)\\ 
\hline \\[-1.8ex] 
 risk aversion $(s)$ & 0.002 &  & $-$0.01 & $-$0.02 & $-$0.03 & $-$0.01 \\ 
  & (0.02) &  & (0.04) & (0.06) & (0.05) & (0.06) \\ 
  & & & & & & \\ 
 mixing intensity $(s)$ &  & 0.01 & 0.03 & 0.06 & 0.05 & $-$0.02 \\ 
  &  & (0.07) & (0.04) & (0.05) & (0.05) & (0.05) \\ 
  & & & & & & \\ 
 female & 0.06 & 0.50$$ & $-$0.002 & $-$0.07 & 0.05 & $-$0.08 \\ 
  & (0.05) & (0.14) & (0.09) & (0.13) & (0.12) & (0.12) \\ 
  & & & & & & \\ 
 experience $(s)$ & 0.04 & $-$0.04 & $-$0.10$$ & $-$0.04 & $-$0.14$$ & $-$0.01 \\ 
  & (0.02) & (0.07) & (0.04) & (0.06) & (0.05) & (0.06) \\ 
  & & & & & & \\ 
 Constant & 0.63$$ & 0.48$$ & 0.15$$ & 0.43$$ & 0.29$$ & 0.32$$ \\ 
  & (0.03) & (0.10) & (0.06) & (0.08) & (0.07) & (0.08) \\ 
  & & & & & & \\ 
\hline \\[-1.8ex] 
Observations & 88 & 88 & 88 & 88 & 88 & 77 \\ 
R$^{2}$ & 0.03 & 0.17 & 0.07 & 0.02 & 0.10 & 0.01 \\ 
\hline 
\hline \\[-1.8ex] 
\end{tabular} 

	\caption{\textbf{Least squares regressions on proxies of attitude and perceptions.} Covariates with marked with (s) were standardized. The standard errors are denoted in parentheses. Mixing intensity is averaged across all domains. The social domain was not recorded for one session due to a technical failure, which results in a lower sample size of 77. Experience denotes a score constructed from the following survey questions: (i) How much experience do you have with probability theory and statistics? (Response scale: 1-7.) (ii) How much experience do you have with gambling? (Response scale: 1-7.) (iii) Did you attend a university lecture on statistics or probability theory? (iv) On average how many even numbers do you expect if throwing a six-sided dice a thousand times?}
	\label{tab:ind}
\end{table}

The summaries for each domain above lend some validation to the experimental procedure. Next, let us consider the more challenging task of analyzing individual-specific measurements. 
Only 4 participants and 8\% of domain-specific choices exhibit a mixing intensity of 1. Thus, most mixing choices are more in line with variational and second-order preferences, contradicting maxmin preferences, as well as $\alpha$-maxmin (Supplementary Section \ref{sec:app_alphamaxmin}) or any other biseparable preferences (Supplementary Section \ref{sec:app_bisep}).

A participant-specific attitude for mixing intensity is supported by the variation in the mixing intensity. A regression with participant fixed effects explains 72\% of this variation (adjusted $R^2 = 0.46$), whereas domain fixed effects explain only 1\% (adjusted $R^2$  $<0$). The same regressions with multiple mixing and the midpoint of the mixing interval as dependent variables, show lower $R^2$ values for the participant fixed effects regressions and domain fixed effects that are different from zero. The pattern is consistent with the interpretation of those indices as ambiguity and probability perception. See Supplementary Section \ref{sec:sup_co} for the full regression results.

Table \ref{tab:ind} shows regression results with the average mixing intensity in column (1) and risk aversion in column (2) as the dependent variable. No correlation between the two can be detected, suggesting that risk aversion and ambiguity aversion are empirically distinct features. Whereas female participants showed more risk aversion, no difference in the mixing intensity (measuring ambiguity aversion) is detectable. Columns (3) to (6) have as the dependent variable the occurrence of multiple mixing (equivalent to a mixing interval with positive length) and are based on the discrete elicitation\footnote{The discrete elicitation is simpler and was elicited first, which reduces measurement error. Further, multiple mixing generated by the discrete elicitation is independent of simple measurement error in the continuous elicitation that was used to construct the mixing intensity.}. Note that Table \ref{tab:ind} takes the indices as given, abstracting from measurement error. See Section \ref{sec:robust} for a more sophisticated approach with stochastic choice. In column (3) to (6), neither risk aversion nor mixing intensity is consistently correlated with multiple mixing. This suggests that mixing intensity and the set of mixing values $q$ measure distinct aspects of ambiguity-sensitive behavior.

An additional control in Table \ref{tab:ind} is \emph{experience} which measures exposure to probability, statistics, and gambling. Experience is correlated with a reduced likelihood of multiple mixing for the risk and stock domain. One standard deviation of the score reduces the likelihood of multiple mixing from  $15\%$ to $5\%$ in the risk domain and from $29\%$ to $15\%$ for the stock index. One possible interpretation is that even sources of uncertainty with clearly defined probabilities can be perceived as ambiguous if participants lack the knowledge and experience. For the ambiguous draw and the social domain, the experience index shows no strong effect, which suggests that ambiguity aversion in those domains is more profound than just a lack of experience with uncertainty.

\subsection{The Prisoner's Dilemma}

Finally, let us consider prosocial behavior in the prisoner's dilemma. 
Individual cooperation is predicted by the midpoint of the belief interval (representing the ambiguity-neutral probability) with a $p$-value $<0.001$ and an estimated effect size of $1\%$ probability of expecting cooperation increasing the likelihood of cooperation by $1.5\%$.  This suggests reciprocity as the main driver of cooperation. For the remaining covariates (risk aversion, mixing intensity, length of the mixing interval) there is no evidence of an effect. The full regression results can be found in the supplementary document. Interestingly, this evidence contradicts previous studies on cooperation and ambiguity that relied on measuring ambiguity with artificial events unrelated to the social game and found a positive correlation between ambiguity tolerance and cooperation \citep{vives2018tolerance}. This simple example illustrates that the elicitation of subjective ambiguity for natural events can sharpen economic analysis and provide new insights.

\section{Robustness}\label{sec:robust}

The theoretical part of the paper derives optimal choices under strong assumptions that thus far guided the analysis. In empirical studies, individuals frequently deviate from this idealized behavior. This section discusses the relevance of \red{hedging, noise, probability weighting, heterogeneity, and aversion to compound objects} for mixing bets. A structural model quantifies the impact of said deviations and adjusts estimates of ambiguity attitude and perception.

\subsection{Hedging}
The theoretical part considered the best response to a single mixing bet in isolation. The experiment elicited multiple mixing bets and tried to induce isolation by presenting decisions on separate screens and paying out only one randomly selected decision. However, there is some evidence that this approach does not universally induce isolation \citep{trautmann2015ambiguity,baillon2021randomize}. If participants integrate across multiple decisions, hedging reduces ambiguity \citep{bade2015randomization}. 

Consider an agent that integrates over mixing bets with a uniform randomization distribution for the values $q \in [0,1]$. The integrating agent can choose $x(q)$ and considers acts
$$l_{x(q)}(s) = [\int \ind{s=E} q x(q) + \ind{s=E^c} (1-q) (1-x(q)) dq]^w_0.$$
If we impose non-mixing and  a single switching point $c$ such that $x(q)=\ind{q>c}$, we obtain 
$$l_{x(q)}(s) = [\ind{s=E} \tfrac{1}{2}(1-c^2) + \ind{s=E^c} \tfrac{1}{2}(1-(1-c)^2)]^w_0,$$
which equals exactly the binarized quadratic scoring rule \citep{brier1950verification,hossain2013binarized}. If the agent always bets on the event with higher odds, i.e., $x(q)=\ind{q>0.5}$, we obtain the lottery $l_{x(q)}(s) = [\frac{3}{8}]^w_0$, i.e., a full hedge by integration.\footnote{I thank an anonymous referee for pointing out the connection to the quadratic scoring rule.} Thus, a fully hedging participant behaves like an ambiguity-neutral agent with belief $p=0.5$. 

The ambiguous urn draw in the experiment has a symmetry point at probability $\tfrac{1}{3}$. Panel A in Figure \ref{fig:exp_averages} indicates an average midpoint of the belief interval that is slightly drawn towards $0.5$ (bias = $0.06$) consistent with some participants hedging. In total, only 3 participants exhibited responses consistent with full hedging ($x(q)=\ind{q>0.5}$ for $q \neq 0.5$), which suggests that full hedging is not common. 
The structural model allows for utility premia for responses consistent with integration to test for moderate hedging.

\subsection{Stochastic Choices}

Choices can be stochastic because of deliberate randomization \citep{agranov2022revealed}, noise/measurement error \citep{gillen2019experimenting}, or random preferences \citep{mcfadden1974conditional}.

The analysis of choices under uncertainty often takes indices as given as in Section \ref{sec:indices} \citep[compare also][]{wakker2018measuring,li2018trust,bleichrodt2023testing,dimmock2015ambiguity,haridon2018off} or assumes independent noise for indices constructed from multiple choices \citep[e.g., on matching probabilities or certainty equivalents, compare][]{baillon2019testing,bruhin2010risk,haridon2019all}. A more granular analysis models each individual choice with a stochastic choice model \citep{gaudecker2011heterogeneity,gao2023behavioral,stahl2014heterogeneity}. For mixing bets, the latter approach avoids the assumption of a parametric error for the length of the mixing interval or the mixing intensity, which are bounded and therefore cannot be modeled adequately with a Gaussian error. 
Instead, I model individual choice probabilities as monotone functions of utilities with a logit model \citep{mcfadden1974conditional} that quantifies the amount of noise in observed choices.

\subsection{Probability Weighting}
The theoretical analysis builds on the Anscombe-Aumann framework, implicitly assuming expected utility for lotteries. In experimental studies, however, many participants showcase insensitivity consistent with probability weighting \citep{quiggin1981risk}, where a lottery $[p]$ is valued with an inversely S-shaped weighting function $W(p)$ \citep{gonzalez1999shape,armantier2016rich}.

Under probability weighting, the preference functional can be non-linear in the winning probability, invalidating the lottery procedure and biasing measurements based on mixing bets. Empirical studies find mixed evidence for the ability of the lottery procedure to improve elicitation \citep[compare for example][]{selten1999money,harrison2015reduction,hossain2013binarized,cox2015paradoxes}.
Under the typical inverse $S$-shape, the weighting function is concave for low probabilities and convex for high probabilities. 
For example, if $x=1$ the two resulting lotteries are $[q]$ and $[0]$, which can be fully in the concave part of probability weighting such that mixing can be preferred, e.g., $x=\tfrac{1}{2}$ resulting in lotteries $[\tfrac{q}{2}]$ and $[\tfrac{1-q}{2}]$. Thus, insensitivity can lead to mixing.

Two measures can be undertaken to improve elicitation under non-linear probability weighting. First, the odds can be rescaled to avoid non-linearities. I show that avoiding certainty and impossibility effects makes the theoretical results of Section \ref{sec:theory} robust to neo-additive weighting functions \citep{chateauneuf2007choice} (see Supplementary Section \ref{sec:pw}, where I build on \cite{baillon2019testing}).

For more general weighting functions, rescaling can reduce the bias, but analytical solutions are beyond the scope of the paper. Instead, I propose to adjust measurements for probability weighting. Following \citet{baillon2019testing}, the structural model considers preferences where the expected utility for lotteries is substituted by a weighted expected utility using the popular weighting function introduced in \citet{prelec1998probability} with insensitivity based on estimates from \citet{haridon2019all}. For an illustration of the effect of probability weighting see Supplementary Section \ref{sec:supp_weighting}.
In the experiment, the odds are scaled such that betting on the event with the higher odds generates a winning chance of $1$. Thus, the weighting function is not strictly concave and ambiguity is not necessarily over-estimated. 
The structural model allows to estimate to which degree choices resemble the patterns expected under different degrees of insensitivity.

\subsection{Aversion to Compound Objects}

Mixing bets are compound objects that depend on the event and some lottery. In empirical studies, participants often prefer a simple lottery to a compound lottery that reduces to the same probability level. This aversion to compound lotteries was found to be linked to ambiguity attitude \citep{halevy2007ellsberg,abdellaoui2015experiments}. That raises the question, if aversion to compound objects could distort the elicitation of ambiguity attitudes with mixing bets. It is worth noting that the mixing choice $x(q)=1-q$ does not result in a compound object. Further, the rescaling of odds in the experiment implies that $x(q) = 1$ for $q\ge\tfrac{1}{2}$ (and respectively $x(q)=0$ for $q\le \tfrac{1}{2}$) also does not constitute a compound object (as the lottery is scaled to certainty). The commonly observed aversion to compound objects would be consistent with an increased likelihood of those choices. In the stochastic choice model, this allows to identify a utility difference for compound objects.

\subsection{Heterogeneity}
Heterogeneity in preferences is important to explain choices under risk \citep{gaudecker2011heterogeneity,bruhin2010risk}. Commonly, beliefs and ambiguity perception have no clear rational benchmark, which renders individual-specific estimates even more relevant \citep{stahl2014heterogeneity}. Further, participant-specific noise has been found to be relevant for decisions under ambiguity \citep{gaudecker2022distribution}. In the structural model, all parameters are estimated at the individual level with a hierarchical model \citep{gao2023behavioral} which allows inference on the distribution of parameters in the population.

\subsection{A Structural Model}

This section proposes a discrete choice model in logit form \citep{mcfadden1974conditional} that extends the deterministic models from Section \ref{sec:theory} along the dimensions discussed above.

The preference basis is captured by variational preferences with
$$U(x,q,B,\theta)= \min_{p \in \bi} E_\gamma(p,x,q) + c_\theta(p),$$
where $c_\theta(p)=\theta K(p||\tfrac{1}{2})$ uses the Kullbeck Leibler divergence and $E_\gamma(p,x,q)$ denotes the rank dependent utility \citep{quiggin1981risk} of a mixing bet with event probability $p$, i.e., 
$$E_\gamma(p,x,q) = p W_\gamma(xq) + (1-p) W_\gamma((1-x)(1-q)),$$
where $W_\gamma(p)=e^{-(-ln(p))^{\gamma}}$ denotes the popular weighting function proposed by \citet{prelec1998probability} such that $\gamma = 1$ recovers expected utility for risk.\footnote{The results shown here were robust to including a second parameter, which is omitted in the main specification.}

The parametric representation additionally allows for utility differences for compound acts ($\beta_c$) and for choices consistent with hedging ($\beta_h$). We define
$$V(x,q,B,\theta,\beta)= U(x,q,\bi,\theta) + \beta_c z_c + \beta_h z_h,$$
where $z_c$ is dummy variable for the resulting mixing bet being a compound object and $z_h$ the product of the length of the belief interval $B$ with a dummy for the response being consistent with hedging. Note that $\beta_c=\beta_h=0$ recovers variational preferences with probability weighting for risk. Finally, for the discrete choice $x \in K$, where $K$ is a finite set in the unit interval, we assume a logit specification where choice probabilities are
$$\dP(x=k) \disas e^{\sigma (V(k,q,B,\theta,\beta) + \epsilon)} \text{ for all }k \in K,$$
consistent with a random utility model with independent, extreme value distributed error terms that scale with $\tfrac{1}{\sigma}$ \citep[Chapter 3,][]{train2009discrete}. For $\sigma \to 0$, responses are fully random. For $\sigma \to \infty$, the choices are deterministic and can recover the optimal responses to mixing bets as derived in Section \ref{sec:theory}. Note that the model assumes independence of error terms only for noise beyond the fixed effects $\epsilon$, which will be used to model dependence between responses of one participant. 

\subsection{Bayesian Estimation}
Flexible discrete choice models can be estimated conveniently with a Bayesian Markov chain Monte Carlo (MCMC) procedure \citep[][Chapter 12]{train2009discrete}. The Bayesian approach has recently gained prominence in the analysis of choice under uncertainty in economic experiments as it allows principled inference on individual specific parameters, using information from all participants in an hierarchical model \citep[e.g.,][]{nilsson2011hierarchical,stahl2014heterogeneity, georgalos2023higher,gao2023behavioral}. The posterior distribution of hyper-parameters can be interpreted as illustrating the representative agent model, while individual-specific posteriors allow to classify participants (e.g., as maxmin type). Bayesian inference remains valid under weak identification, which can arise as both the absence of ambiguity aversion and perception are consistent with non-mixing.

The belief interval is parameterized by a midpoint $m$ and length of the belief interval $l$ such that $B=[m-\tfrac{1}{2}l,m+\tfrac{1}{2}l]$. The full parameter vector is given by $(m_{dp},l_{dp}, \theta_p, \gamma_p,\beta_p,\sigma_p,\epsilon_p)$, where subscripts $p$ denote the participant and $d$ the domain. Note that $\epsilon_p$ is vector with 11 entries, which captures the participant-specific preferences for choices $x\in\{0,0.1,\dots,1\}$.

To improve computational efficiency, the parameters influencing the utility $U$ are defined on a discrete grid, which allows to compute $U(x,q,B,\theta)$ for all parameter values in parallel before the MCMC procedure. Conveniently, posterior means can still attain all values as averages over MCMC draws. An additional advantage is that the discrete distribution on a grid avoids shape restrictions and can approximate any distribution.

To estimate individual-specific parameters, a hierarchical model is used that allows high flexibility, while using the information across participants to shrink parameters towards a shared representative agent value \citep[see][and references therein]{gao2023behavioral}. 

For the parameters $\beta_p$ and $\sigma_p$ normal and half-normal priors are assumed. The same is true for the hyper-parameters that have almost flat priors. For the remaining parameters, defined on the grid, categorical distributions with Dirichlet hyper-priors are used. (See Supplementary Section \ref{sec:priors} for details.)

\subsection{Results}

\begin{figure}
	\includegraphics[width=\textwidth]{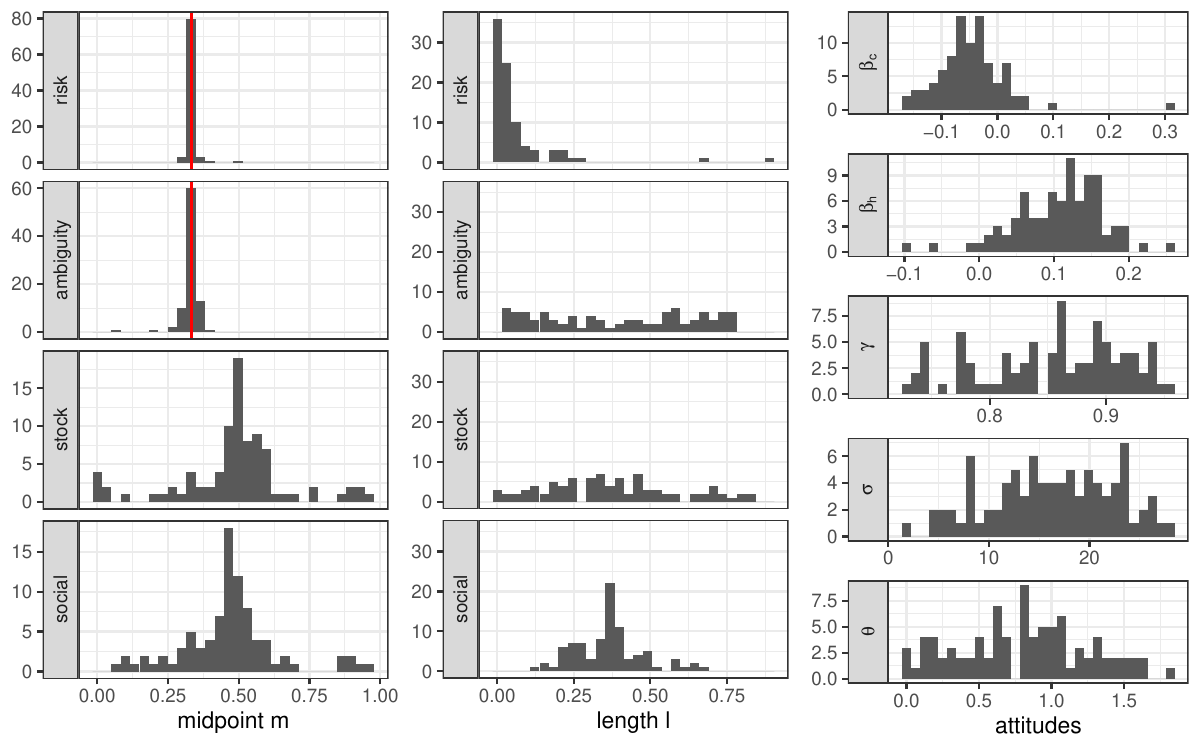}
	\caption{\textbf{Distribution of posterior means across participants.} The plot depicts histograms of posterior means of all participants. The first column shows the midpoint $m$ of the belief interval (the red line indicates the symmetry point of $\tfrac{1}{3}$ for the urn draws), the second column the length $l$ of the belief interval, and the third column the parameters for compound objects $\beta_c$, hedging choices $\beta_h$, insensitivity in probability weighting $\gamma$, noise $\sigma$, and ambiguity attitude $\theta$.}
	\label{fig:rum_individual}
\end{figure}

The analysis focuses on the distribution of parameters across participants. Figure \ref{fig:rum_individual} shows the distribution of posterior mean estimates for all parameters. (Compare Table \ref{tab:rum_ind} in Supplementary Section \ref{sec:rum} for additional results).
%
The first row shows that the midpoint of the belief interval is estimated to be close to $\tfrac{1}{3}$ for the risky and ambiguous draw for all participants. The beliefs on the two natural events are much more dispersed (especially for the stock index). 

In the second column of Figure \ref{fig:rum_individual}, the length of the belief interval is estimated to be small for most participants for the risky draw. Specifically, the average posterior probability for $l_{p,risk}=0$ is $91\%$ and the posterior mode of $98\%$ of participants is $0$.
For the ambiguous domain 44\% of posterior modes are strictly positive and the average posterior probability of $l_{p,ambiguity}=0$ is only $41\%$. 
The average estimate for the length of the belief interval for the natural events is closer to the ambiguous draw than the risky draw ($0.36$ for social, $0.37$ for stock, $0.40$ for the ambiguous and $0.06$ for the risky draw). A striking difference arises in the spread of ambiguity perception, with participants differing widely in their ambiguity perception for the stock index but comparatively little for the social domain. 

While there remains substantial uncertainty in the posterior distributions of participant-specific ambiguity estimates, on population-level the three potentially ambiguous domains show strong evidence of being perceived as more ambiguous than the risky draw with a posterior probability of $> 99.99\%$ that the hyper-parameter mean of the length of the belief interval is larger. 

The model estimates moderate levels of aversion to compound objects with an average utility difference of $-0.05$ (standard utilities range from $0$ and $1$). The posterior probability of participants being averse to compound objects on average is $99.6\%$. Significant uncertainty remains for the individual estimates, with only $27\%$ of participants having $90\%$-credible sets that do not contain the zero. 
A similar effect is estimated for hedging choices, where the average utility difference is $0.11$ multiplied with the length of the belief interval (on average $0.4$ for ambiguity). The posterior probability of participants having a preference for hedging on average is $76.2\%$. Individual estimates were too noisy to reject a zero effect for any participant based on the $90\%$-credible sets. As comparison, \citet{baillon2021randomize} found that about half of all participants showed signs of hedging.

Results are consistent with moderate levels of insensitivity. The posterior mean estimates of $\gamma_p$ cover the range from $0.7$ to $1.0$. \citet{haridon2019all} estimate a population mean of $0.75$ for a German sample. Such insensitivity would imply less mixing for symmetric odds and more for asymmetric odds compared to no probability weighting (Supplementary Section \ref{sec:supp_weighting}).

In line with the literature \citep{stahl2014heterogeneity,gaudecker2022distribution}, the model reveals substantial deviations from deterministic choice. The posterior mean estimates of $\sigma_p$ range from $2$ to $29$. The lowest values are consistent with essentially random answers. For $\sigma_p=10$, about $80\%$ of choices from the discrete elicitation coincide with their deterministic version, and the mean absolute error for the continuous elicitation is $0.2$. For $\sigma_p=30$, about $90\%$ of choices from the discrete elicitation coincide with their deterministic version, and the mean absolute error for the continuous elicitation is below $0.1$ (see Supplementary Section \ref{sec:choice_prob}).

An ambiguity attitude of $\theta=0$ is consistent with maxmin preferences. The average posterior probability for $\theta_p=0$ is $51\%$, consistent with about half of the ambiguity-averse participants being maxmin type.

To sum up, the structural model found strong evidence for noise, and some evidence for hedging, probability weighting, and aversion to compound objects. This suggests that testing for ambiguity has to take into account those factors. The main results of the index-based analysis are robust: Mixing bets can reveal ambiguity, and participants perceived almost as much ambiguity for the natural events as for the Ellsberg urn.

\section{Discussion}\label{sec:discussion}

The separation of attitude and perception is a potentially insightful endeavor \citep{manski2004measuring,bleichrodt2023testing}. While the ability of ambiguity-sensitive models to achieve said separation is a matter of ongoing debate \citep{epstein2010paradox,klibanoff2012smooth}, mixing bets can be used to elicit several different aspects of subjective ambiguity. 
In particular, I propose three indices that can be understood as measuring the ambiguity-neutral probability, ambiguity perception, and ambiguity attitude. This interpretation is most adequate for variational preferences that contain maxmin preferences as limiting case. For second-order preferences, the separation is more challenging as the length of the mixing interval for small stakes is dominated by an interaction term between the second-order utility function and the belief distribution. 
In an experimental study, I find that the three indices and risk aversion are uncorrelated, which supports their interpretation as different aspects of subjective uncertainty. 
I show that mixing bets can help explain economic decisions. In particular, I find that the midpoint of the mixing interval (measuring probability), but not its length or the mixing intensity, correlates with cooperation in a social game.
Further, I show that mixing bets can be used to analyze the change of ambiguity perception under new information. In particular, additional draws from the Ellsberg urn reduced the extent of mixing.

I discuss the relevance of behavioral phenomena for eliciting ambiguity with mixing bets. Noise, probability weighting, and aversion to compound objects can increase mixing, while hedging can decrease mixing. A structural model is proposed that finds evidence for all four phenomena, where estimates are consistent with considerable noise and moderate levels of probability weighting, aversion to compound objects, and hedging. Yet, the structural model confirms the findings of a simple index-based analysis: Ambiguity perception for the stock market and the social domain are almost as high as for the Ellsberg urn, which underlines the relevance of ambiguity in economic decision-making. Finally, ambiguity and probability perception were heterogeneous, which could explain the strong variation between individuals in an international comparison of ambiguity attitudes \citep{haridon2018off}.

The main alternative to mixing bets are belief hedges \citep{wakker2018measuring,baillon2018balanced}. Belief hedges require at least three events to generate the insensitivity index that is often argued to capture ambiguity perception. Similar to mixing bets, they do not achieve a full separation for second-order preferences \citep[see Observation~18][]{baillon2018balanced}. However, belief hedges can separate ambiguity attitude and perception for the neo-additive model of \citet{chateauneuf2007choice} \citep[see also][]{baillon2018effect}.

\red{Belief hedges are based on \emph{matching probabilities} \citep[e.g.,][]{holt2007markets} also called \emph{choice-based probabilities} \citep{abdellaoui2011rich}.  
Conveniently, matching probabilities can be analyzed without a mixing concept or a product state space, whereas mixing bets are compound objects between a lottery and the analyzed event, which makes measurement vulnerable to aversion to compound objects. Further, it is more complicated to disentangle probability weighting on lotteries from ambiguity attitudes, where the literature on matching probabilities already provides a wealth of approaches \citep{baillon2019testing,abdellaoui2011rich,wakker2018measuring,baillon2018balanced}.
Comfortably, the experimental evidence for the Ellsberg urn provided here is broadly in line with a long line of experiments based on matching probabilities: On average participants avoided ambiguous uncertainty more than pure risk.
Future research might be able to connect ideas from belief hedges and mixing bets. In applications, it can be beneficial to elicit both to adjust for measurement error by multiple measurements \citep{gillen2019experimenting}.}

The experiment focused on the measurement of ambiguity aversion. If, additionally, ambiguity seeking should be explored, separated mixing bets can distinguish ambiguity-seeking from ambiguity-averse preferences and bound the belief interval irrespective of ambiguity attitude. However, for ambiguity-seeking preferences the mixing intensity is not observable. 
\red{Non-uniform ambiguity attitude \citep[see][for empirical evidence]{baillon2015testing}, where preferences are neither ambiguity-averse nor ambiguity-seeking, are beyond the scope of the paper. For $\alpha$-maxmin preferences mixing bets require an additional step to fully reveal ambiguity (compare Supplementary Section \ref{sec:app_alphamaxmin}). For other recent contributions on the elicitation of $\alpha$-maxmin preferences see \citet{hill2023beyond,hill2021eliciting,bleichrodt2023testing,bose2017eliciting}.}

A major concern is how the agent reacts when faced with multiple bets on the same uncertain outcome \citep{bade2015randomization}. The elicitation of ambiguity requires the agent to consider each choice separately. \citet{baillon2021randomize} find that about half of their ambiguity-averse participants show evidence of hedging. For mixing bets, hedging was found to be of minor influence, which suggests that its relevance differs between populations or settings. If group-based inference is sufficient, robustness to hedging can be achieved by eliciting just a single choice from each participant.

Mixing bets have a direct connection to proper scoring rules \citep{Gneiting} and can be seen as an application of multiple quantile elicitation as introduced in \citet{eyting2021belief} to binary events. Mixing bets can also be connected to early work on subjective probabilities. Preceding ambiguity-sensitive decision models, \citet{smith1961consistency} proposed to define subjective probabilities by the odds that an agent agrees to bet on an event. By allowing the agent to mix, I provide a generalization of Smith's hypothetical design and establish that mixing identifies ambiguity.

Instead of using mixing bets, one can ask participants directly for ranges of probabilities \citep[e.g.,][]{manski2010rounding,giustinelli2021precise}. However, ambiguity-averse decision models describe behavior, rather than thought processes. The belief interval may well have considerable explanatory power regarding an agent's behavior, while the agent is unable or unwilling to articulate such an interval. In this case, a revealed preferences approach is more suitable.

\appendix
\section*{Appendix: Proofs}

\begin{proof}[Proof of Lemma \ref{th:betseu}]
	Under the Anscombe-Aumann framework the expected utility is simply
	$$\dE_{s \disas p}[u(l(s))] = p u([xq]^w_0) + (1-p) u([(1-x)(1-q)]^w_0),$$ 
	assuming independence between the event $E$ and the lotteries. Normalizing $u(0)=0$ and $\du = u(w)-u(0)$, we have $u([xq]^w_0)= xq \du$ and $u([(1-x)(1-q)]^w_0) = (1-x)(1-q) \du$, which gives
	$$\dE_{s \disas p}[u(l(s))] = (1 - x - p - q + xp + xq + pq) \du = (x (p - (1-q)) + pq + 1) \du.$$
\end{proof}

\begin{proof}[Proof of Lemma \ref{th:betsvariationallimited}]
	We define $s(q,x,p) \defeq 1 - x - p - q + xp + xq + pq$, such that Lemma \ref{th:betseu} implies 
	$\dE_{s \disas p}[u(l(s))] = s(q,x,p) \du$, 
	and we obtain the simplified optimization problem	
	$$x^\ast(q) = \argmax_{x \in [0,1]} \; \;
	\min_{p \in \bi} \; s(q,x,q) + c(p)/\du.$$
	The decision-maker acts as if more ambiguity averse for higher utility difference between prizes $\du$ \citep[Proposition 8,][]{maccheroni2006ambiguity}. Define $c_t(p)=c(p)/\du$ for notational convenience. $c_t$ is also grounded, strictly convex and 
	twice continuously differentiable.
	As $c_t$ is convex it follows that $c_t'$ is increasing and as $c_t$ is grounded, it follows that $c'_t(a)\le0$ and $c'_t(b)\ge0$.
	
	Next, we investigate $x^\ast(q)$ for fixed $q\in[0,1]$. Examine the minimum of
	\begin{align}
		g_x(p) \defeq s(q,x,q) + c_t(p) = 1 - x - q + xq + p(x-(1-q)) + c_t(p).
	\end{align}
	The function $g_x$ is convex.	The minimum at $p^\ast(x)\defeq \argmin_{p\in B} g_x(p)$ is characterized by $g'_x(p)=x-(1-q) + c'_t(p)$. 
	
	\begin{itemize}
		\item First case: $\pstar=\av$.\\
		In this case, $g'_x(\pstar)\ge0$ or equivalently $x\ge1-q-c'_t(\av)$.
		Further, the agent values the resulting bets as a function of $x$ by
		\begin{align}
			U(x)&= g_x(a)= 1 - \av - q + \av q + x (\av-(1-q)) + c_t(\av).
		\end{align}
		If $1-q<\av$, it follows that $x^\ast=1$. If $1-q>\av$, the smallest feasible value is optimal and $x^\ast=\min(1,1-q-c'_t(\av)).$
		
		\item Second case: $\pstar=\bv$.\\
		In this case, $g'_x(\pstar)\le 0$ or equivalently $x\le 1-q-c'_t(b)$. 
		Further, the agent values the resulting bets as a function of $x$ by
		\begin{align}
			U(x) = 1 - b - q + \bv q + x (\bv-(1-q)) + c_t(\bv).
		\end{align}
		If $1-q>\bv$, it follows that $x^\ast=0$. For $1-q<\bv$, the largest feasible value is optimal and $x^\ast=\max(0,1-q-c'_t(\bv)).$
				
		\item Third case: $\pstar \in (\av,\bv)$.
		In this case $g'_x(\pstar)=0$ or equivalently $x=1-q-c'_t(\pstar)$.
	
		As $c_t'$ is increasing, it follows that $\pstar$ is decreasing in $x$.	
		The agent values the resulting bets as a function of $x$ by
		\begin{align}
			U(x) = 1 - x - q + xq + \pstar (x-(1-q)) + c_t(\pstar).
		\end{align}

		The first order condition is
		\[
		\begin{alignedat}{3}
			- (1 - q) + \pstar' x + \pstar - \pstar' (1-q) + c'_t(\pstar) \pstar'&=0,\\
			\pstar + \pstar' (x-(1-q)) + c'_t(\pstar) \pstar' &= 1 - q,\\
			\pstar - \pstar' c'_t(\pstar) + c'_t(\pstar) \pstar' &= 1 - q,\\
			\pstar &=  1 - q,
		\end{alignedat}
		\]
		and describes a maximum as $U'(x)>0 \iff \pstar > 1-q$ and $\pstar$ decreasing in $x$. 
		Thus, it follows that 
		\[
		x^\ast(q)= 
		\left\{\begin{alignedat}{3}
			&0 && if \; c'_t(1-q) \ge &&1-q,\\
			&1-q - c'_t(1-q) \quad && if \; c'_t(1-q) > &&1-q>1+c'_t(1-q),\\
			&1 && if \; &&1-q \ge 1+ c'_t(1-q).
		\end{alignedat}\right.
		\]
		If $c(1-q)=0$, it follows that $c'(1-q)=0$ as $c$ is grounded and convex and thus $x^\ast(q)=1-q$. 
		For any $q\in B$, mixing is optimal if 
		\begin{align}
			\frac{c'(1-q)}{\du}&< \; 1-q \; <1+\frac{c'(1-q)}{\du},
		\end{align}
		which holds true for a sufficiently large $\du$ if $c'$ is bounded.
	\end{itemize}
\end{proof}

\begin{proof}[Proof of Lemma \ref{th:betssmooth}]
	With Lemma \ref{th:betseu} it holds that
	$$x^\ast(q) = \argmax_{x \in [0,1]} \; \;
	\; \dE_{p \disas \dP}[\, \phi_t(s(q,x,q)) \,],$$
	with $\phi_t(z)=\phi(\du z + u(0))$ increasing and concave and $s(q,x,q)= 1 - p - q + pq + x(p-(1-q))$. 
	
	First, consider the case $1-q \le a$. As $p \le 1-q$ implies $\phi_t$ is increasing in $x$, this case implies that $\phi_t$ is $\dP$-almost surely increasing in $x$. Thus, $\dE_{p \disas \dP}[\, \phi_t(s(q,x,q)) \,]$ increasing in $x$ and $x^\ast=1$. A similar argument shows $x^\ast=0$ for $1-q\ge b$. 
	
	The remainder of the proof considers the case $a<1-q<b$. Let $U(x,q)=\dE_{p \disas \dP}[\, \phi_t(s(q,x,q)) \,].$ As $\phi_t$ is continuously differentiable, $\phi_t$ and its first two derivatives are integrable on $\bi$, it follows by the dominant convergence theorem that $(\partial_x)^2 U(x,q)=
	\dE_{p \disas \dP}[ \phi''_t(s(q,x,q)) (p-(1-q))^2],$
	which in turn implies that $U(x,q)$ is concave in $x$ as $\phi''_t \le 0$. We conclude that for fixed $q$ the optimal mixing $x^\ast(q)$ is unique.
	Further, by the maximum theorem \citep{Ok2007} $x^\ast(q)$ is continuous as it holds that $U(x,q)$ is continuous by the dominated convergence theorem.
	
	If $a\neq b$ the following argument shows that mixing is optimal for an interval that contains $1-\dE_{p \disas \dP}[p]$. Consider the first order condition $\partial_x U(x,q)=\dE_{p \disas \dP}[\phi_t'(s(q,x,q)) (p-(1-q))]=0.$ For $x=1$, the equation above is equivalent to
	$\dE_{p \disas \dP}[\phi_t'(pq) (p-(1-q))]=0.$ As $\phi_t$ concave, the derivative $\phi_t'$ is decreasing and it follows that $\phi_t'(pq)\le \phi'_t(b q)$ almost surely. Thus,
	$\dE_{p \disas \dP}[\phi_t'(pq) (p-(1-q))] \le \phi_t'(bq) (\dE_{p \disas \dP}[p]-(1-q))<0,$
	for $1-q>\dE_{p \disas \dP}[p]$. Analogously, it can be followed that the FOC for $x=0$ is positive if $1-q<\dE_{E \disas p}[p]$. As $x^\ast(1-q)$ is continuous on the belief interval $\bi$, it follows that mixing is optimal in an environment of $\dE_{E \disas p}[p]$ if $\bi$ doesn't reduce to a single point.
	
	Now consider a series $w_n$ such that $u_{\Delta,n}=u(w_n)-u(0) \to \infty$. The utility function is not unique \citep[compare Theorem 1][]{klibanoff2005smooth}. If preferences are represented by utility functions $u_n(0)=0$ and $u_n(w_n)=1$, \red{each $w_n$ corresponds to preferences with transformed $\phi_n(z)=\phi(\tfrac{z}{u_{\Delta,n}} + c_0)$. Next, we consider this series of preferences with their respective coefficients of ambiguity aversion denoted by}
	$$\alpha_n(z)=-\frac{\phi_n''}{\phi_n'}=-\frac{u_{\Delta,n} \phi''(u_{\Delta,n} z + u(0))}{u_{\Delta,n}^2 \phi'(u_{\Delta,n} z + u(0))}=\frac{\alpha(z)}{u_{\Delta,n}},$$
	where $\alpha(z)=-\tfrac{\phi''(z)}{\phi'(z)}$ is the uniquely defined coefficient of ambiguity aversion of the initial preferences. All considered preferences share the same second order beliefs $\dP$ with support $B$. It holds that $\alpha_{n+1}>\alpha_{n}$. If $\alpha(z)$ is bounded away from zero, it holds that $\inf_z \alpha_n(z) \to \infty$. \red{With Proposition 3 in \citet{klibanoff2005smooth} it follows that if a maxmin agent with belief set $B$ strictly prefers mixing for a value of $q$ it holds that for large enough $w_n$ the members of our series also strictly prefer mixing. As maxmin preferences with belief interval $B$ strictly prefer mixing for any $1-q \in (a,b)$, it holds that there exists a $w_n$ such that $1-q \in M_{\du}$.
}
	
\end{proof}

{\renewcommand*{\bibfont}{\small}
\printbibliography

@techreport { schmidt2024replication ,
	author = { Schmidt, Patrick },
	year = {2024} ,
	title = { "Data and code for: Eliciting ambiguity with mixing bets" American Economic Association} ,
	institution = { Inter-university Consortium for Political and Social Research } ,
	type = {} ,
	number = { openicpsr-197461 },
	note = { http://doi.org/10.3886/E202864V1 } ,
}

@unpublished{hill2021eliciting,
	title={Eliciting multiple prior beliefs},
	author={Abdellaoui, Mohammed and Colo, Philippe and Hill, Brian},
	journal={HEC Paris Research Paper No. ECO/SCD-2021-1426},
	year={2021},
	note = {working paper}
}

@article{cox2015paradoxes,
	title={Paradoxes and mechanisms for choice under risk},
	author={Cox, James C and Sadiraj, Vjollca and Schmidt, Ulrich},
	journal={Experimental Economics},
	volume={18},
	pages={215--250},
	year={2015},
	publisher={Springer}
}

@article{selten1999money,
	title={Money does not induce risk neutral behavior, but binary lotteries do even worse},
	author={Selten, Reinhard and Sadrieh, Abdolkarim and Abbink, Klaus},
	journal={Theory and Decision},
	volume={46},
	pages={213--252},
	year={1999},
	publisher={Springer}
}

@article{giustinelli2021precise,
	author = {Giustinelli, Pamela and Manski, Charles F and Molinari, Francesca},
	title = "{Precise or imprecise probabilities? Evidence from survey response related to late-onset dementia}",
	journal = {Journal of the European Economic Association},
	volume = {20},
	number = {1},
	pages = {187-221},
	year = {2021},
	month = {07}
}

@article{nilsson2011hierarchical,
	title={Hierarchical Bayesian parameter estimation for cumulative prospect theory},
	author={Nilsson, H{\aa}kan and Rieskamp, J{\"o}rg and Wagenmakers, Eric-Jan},
	journal={Journal of Mathematical Psychology},
	volume={55},
	number={1},
	pages={84--93},
	year={2011},
	publisher={Elsevier}
}

@article{anantanasuwong2019ambiguity,
	title={Ambiguity attitudes about investements: {Evidence} from the field},
	author={Anantanasuwong, Kanin and Kouwenberg,Roy and Mitchell, Olivia S. and Peijnenberg, Kim},
	year={2019},
	journal={NBER Working Paper 25561}
}

@article{stahl2014heterogeneity,
	title={Heterogeneity of ambiguity preferences},
	author={Stahl, Dale O},
	journal={Review of Economics and Statistics},
	volume={96},
	number={4},
	pages={609--617},
	year={2014},
	publisher={The MIT Press}
}

@article{dominiak2011attitudes,
	title={Attitudes toward uncertainty and randomization: an experimental study},
	author={Dominiak, Adam and Schnedler, Wendelin},
	journal={Economic Theory},
	volume={48},
	number={2-3},
	pages={289--312},
	year={2011},
	publisher={Springer}
}

@article{anscombe1963definition,
	title={A definition of subjective probability},
	author={Anscombe, Francis J and Aumann, Robert J},
	journal={Annals of Mathematical Statistics},
	volume={34},
	number={1},
	pages={199--205},
	year={1963}
}

@article{klibanoff2012smooth,
	title={On the smooth ambiguity model: A reply},
	author={Klibanoff, Peter and Marinacci, Massimo and Mukerji, Sujoy},
	journal={Econometrica},
	volume={80},
	number={3},
	pages={1303--1321},
	year={2012},
	publisher={Wiley Online Library}
}

@article{epstein2010paradox,
	title={A paradox for the “smooth ambiguity” model of preference},
	author={Epstein, Larry G},
	journal={Econometrica},
	volume={78},
	number={6},
	pages={2085--2099},
	year={2010},
	publisher={Wiley Online Library}
}

@article{vives2018tolerance,
	title={Tolerance to ambiguous uncertainty predicts prosocial behavior},
	author={Vives, Marc-Llu{\'\i}s and FeldmanHall, Oriel},
	journal={Nature Communications},
	volume={9},
	number={1},
	pages={1--9},
	year={2018},
	publisher={Nature Publishing Group}
}

@article{eyting2021belief,
	title = {Belief elicitation with multiple point predictions},
	journal = {European Economic Review},
	volume = {135},
	year = {2021},
	issn = {0014-2921},
	pages = {103700},
	doi = {https://doi.org/10.1016/j.euroecorev.2021.103700},
	url = {https://www.sciencedirect.com/science/article/pii/S0014292121000532},
	author = {Markus Eyting and Patrick Schmidt}
}

@article{prelec1998probability,
	title={The probability weighting function},
	author={Prelec, Drazen},
	journal={Econometrica},
	volume={66},
	number={3},	
	pages={497--527},
	year={1998},
	publisher={JSTOR}
}

@article{machina1992more,
	title={A more robust definition of subjective probability},
	author={Machina, Mark J and Schmeidler, David},
	journal={Econometrica},
	number={4},
	volume={60},
	pages={745--780},
	year={1992},
	publisher={JSTOR}
}

@article{marinacci2002probabilistic,
	title={Probabilistic sophistication and multiple priors},
	author={Marinacci, Massimo},
	journal={Econometrica},
	volume={70},
	number={2},
	pages={755--764},
	year={2002},
	publisher={Wiley Online Library}
}

@article{dow1992uncertainty,
	title={Uncertainty aversion, risk aversion, and the optimal choice of portfolio},
	author={Dow, James and Werlang, Sergio Ribeiro},
	journal={Econometrica},
	volume = {60},
	number={1},
	pages={197--204},
	year={1992},
	publisher={JSTOR}
}

@article{dimmock2015ambiguity,
	title={Ambiguity attitudes in a large representative sample},
	author={Dimmock, Stephen and Kouwenberg, Roy and Wakker, Peter},
	journal={Management Science},
	volume={62},
	number={5},
	pages={1363--1380},
	year={2015},
	publisher={INFORMS}
}

@article{cerreia2011uncertainty,
	title={Uncertainty averse preferences},
	author={Cerreia-Vioglio, Simone and Maccheroni, Fabio and Marinacci, Massimo and Montrucchio, Luigi},
	journal={Journal of Economic Theory},
	volume={146},
	number={4},
	pages={1275--1330},
	year={2011},
	publisher={Academic Press}
}

@article{gaudecker2022distribution,
	title={The distribution of ambiguity attitudes},
	author={Gaudecker, Hans-martin Von and Wogrolly, Axel and Zimpelmann, Christian},
	year={2022},
	publisher={CESifo Working Paper}
}

@article{klibanoff2014perceived,
	title={Perceived ambiguity and relevant measures},
	author={Klibanoff, Peter and Mukerji, Sujoy and Seo, Kyoungwon},
	journal={Econometrica},
	volume={82},
	number={5},
	pages={1945--1978},
	year={2014},
	publisher={Wiley Online Library}
}

@article{abdellaoui2015experiments,
	title={Experiments on compound risk in relation to simple risk and to ambiguity},
	author={Abdellaoui, Mohammed and Klibanoff, Peter and Placido, L{\ae}titia},
	journal={Management Science},
	volume={61},
	number={6},
	pages={1306--1322},
	year={2015},
	publisher={INFORMS}
}

@incollection{trautmann2015ambiguity,
	author = {Trautmann, Stefan T and {van de Kuilen}, Gijs},
	publisher = {John Wiley Sons},
	title = {Ambiguity attitudes},
	booktitle = {The Wiley Blackwell Handbook of Judgment and Decision Making},
	editor={Keren, Gideon and Wu, George},
	chapter = {3},
	pages = {89--116},
	year = {2015}
}

@article{halevy2007ellsberg,
	title={Ellsberg revisited: An experimental study},
	author={Halevy, Yoram},
	journal={Econometrica},
	volume={75},
	number={2},
	pages={503--536},
	year={2007},
	publisher={Wiley Online Library}
}

@article{smith1961consistency,
	title={Consistency in statistical inference and decision},
	author={Smith, Cedric},
	journal={Journal of the Royal Statistical Society. Series B (Methodological)},
	pages={1--37},
	year={1961},
	volume={23},
	number={1},
	publisher={JSTOR}
}

@article{schmeidler1989subjective,
	title={Subjective probability and expected utility without additivity},
	author={Schmeidler, David},
	journal={Econometrica},
	pages={571--587},
	year={1989},
	publisher={JSTOR}
}

@article{ghirardato2001risk,
	title={Risk, ambiguity, and the separation of utility and beliefs},
	author={Ghirardato, Paolo and Marinacci, Massimo},
	journal={Mathematics of Operations Research},
	volume={26},
	number={4},
	pages={864--890},
	year={2001},
	publisher={INFORMS}
}

@article{li2018trust,
	title={Trust as a decision under ambiguity},
	author={Li, Chen and Turmunkh, Uyanga and Wakker, Peter},
	journal={Experimental Economics},
	year={2018}
}

@article{baillon2021randomize,
	author={Baillon, Aur{\'e}lien and Halevy, Yoram and Li, Chen},
	title={Randomize at your own risk: On the observability of ambiguity aversion},
	journal = {Econometrica},
	volume = {90},
	number = {3},
	pages = {1085-1107},
	year = {2022}
}

@article{baillon2018balanced,
title = {Belief hedges: Measuring ambiguity for all events and all models},
journal = {Journal of Economic Theory},
volume = {198},
pages = {105353},
year = {2021},
issn = {0022-0531},
doi = {https://doi.org/10.1016/j.jet.2021.105353},
url = {https://www.sciencedirect.com/science/article/pii/S0022053121001708},
author = {Aurélien Baillon and Han Bleichrodt and Chen Li and Peter P. Wakker}
}

@article{hansen2007beliefs,
	title={Beliefs, doubts and learning: Valuing macroeconomic risk},
	author={Hansen, Lars Peter},
	journal={American Economic Review},
	volume={97},
	number={2},
	pages={1--30},
	year={2007}
}

@article{gillen2019experimenting,
	title={Experimenting with measurement error: Techniques with applications to the Caltech cohort study},
	author={Gillen, Ben and Snowberg, Erik and Yariv, Leeat},
	journal={Journal of Political Economy},
	volume={127},
	number={4},
	pages={1826--1863},
	year={2019},
	publisher={The University of Chicago Press Chicago, IL}
}

@article{hansen2007recursive,
	title={Recursive robust estimation and control without commitment},
	author={Hansen, Lars Peter and Sargent, Thomas J},
	journal={Journal of Economic Theory},
	volume={136},
	number={1},
	pages={1--27},
	year={2007},
	publisher={Elsevier}
}

@article{baillon2015testing,
	title={Testing ambiguity models through the measurement of probabilities for gains and losses},
	author={Baillon, Aur{\'e}lien and Bleichrodt, Han},
	journal={American Economic Journal: Microeconomics},
	volume={7},
	number={2},
	pages={77--100},
	year={2015}
}

@article{baillon2019testing,
	title={Testing constant absolute and relative ambiguity aversion},
	author={Baillon, Aur{\'e}lien and Placido, L{\ae}titia},
	journal={Journal of Economic Theory},
	volume={181},
	pages={309--332},
	year={2019},
	publisher={Elsevier}
}

@book{holt2007markets,
	title={Markets, Games, \& Strategic Behavior},
	author={Holt, Charles A},
	year={2007},
	publisher={Pearson Addison Wesley Boston, MA}
}

@article{wakker2018measuring,
	AUTHOR = {Baillon, A. and Huang, Z. and Selim, A. and Wakker, P.},
	TITLE = {Measuring ambiguity attitudes for all (natural) events},
	JOURNAL = {Econometrica},
	VOLUME = {86},
	YEAR = {2018},
	NUMBER = {5},
	PAGES = {1839--1858}}

@article{klibanoff2005smooth,
	title={A smooth model of decision making under ambiguity},
	author={Klibanoff, Peter and Marinacci, Massimo and Mukerji, Sujoy},
	journal={Econometrica},
	volume={73},
	number={6},
	pages={1849--1892},
	year={2005},
	publisher={Wiley Online Library}
}

@article{maccheroni2006ambiguity,
	title={Ambiguity aversion, robustness, and the variational representation of preferences},
	author={Maccheroni, Fabio and Marinacci, Massimo and Rustichini, Aldo},
	journal={Econometrica},
	volume={74},
	number={6},
	pages={1447--1498},
	year={2006},
	publisher={Wiley Online Library}
}

@article{ghirardato2004differentiating,
	title={Differentiating ambiguity and ambiguity attitude},
	author={Ghirardato, Paolo and Maccheroni, Fabio and Marinacci, Massimo},
	journal={Journal of Economic Theory},
	volume={118},
	number={2},
	pages={133--173},
	year={2004},
	publisher={Elsevier}
}

@article{gilboa1989maxmin,
	title={Maxmin expected utility with non-unique prior},
	author={Gilboa, Itzhak and Schmeidler, David},
	journal={Journal of Mathematical Economics},
	volume={18},
	number={2},
	pages={141--153},
	year={1989},
	publisher={Elsevier}
}

@unpublished{henkel2022experimental,
	title={Experimental evidence on the relationship between perceived ambiguity and likelihood insensitivity},
	author={Henkel, Luca},
	year={2022},
	note = {Working paper}
}

@article{bose2017secondorder,
title = {Eliciting second-order beliefs},
journal = {Journal of Mathematical Economics},
volume = {107},
pages = {102865},
year = {2023},
issn = {0304-4068},
doi = {https://doi.org/10.1016/j.jmateco.2023.102865},
author = {Subir Bose and Arup Daripa}}

@article{ellsberg1961risk,
	title={Risk, ambiguity, and the Savage axioms},
	author={Ellsberg, Daniel},
	journal={The Quarterly Journal of Economics},
	pages={643--669},
	volume={75},
	year={1961},
	publisher={JSTOR}
}

@article{bose2017eliciting,
title = {Eliciting ambiguous beliefs using constructed ambiguous acts: Alpha-maxmin},
journal = {Journal of Mathematical Economics},
volume = {103},
pages = {102787},
year = {2022},
issn = {0304-4068},
doi = {https://doi.org/10.1016/j.jmateco.2022.102787},
url = {https://www.sciencedirect.com/science/article/pii/S0304406822001136},
author = {Subir Bose and Arup Daripa}
}

@article{hill2023beyond,
title = {Beyond uncertainty aversion},
journal = {Games and Economic Behavior},
volume = {141},
pages = {196-222},
year = {2023},
author = {Brian Hill}
}

@article{harrison2015reduction,
	title = {Reduction of compound lotteries with objective probabilities: Theory and evidence},
	journal = {Journal of Economic Behavior \& Organization},
	volume = {119},
	pages = {32-55},
	year = {2015},
	issn = {0167-2681},
	doi = {https://doi.org/10.1016/j.jebo.2015.07.012},
	url = {https://www.sciencedirect.com/science/article/pii/S0167268115002036},
	author = {Glenn W. Harrison and Jimmy Martínez-Correa and J. Todd Swarthout}
}

@article{li2018rich,
	title={The rich domain of ambiguity explored},
	author={Li, Zhihua and M{\"u}ller, Julia and Wakker, Peter P and Wang, Tong V},
	journal={Management Science},
	volume={64},
	number={7},
	pages={3227--3240},
	year={2018},
	publisher={INFORMS}
}

@article{abdellaoui2011rich,
	title={The rich domain of uncertainty: Source functions and their experimental implementation},
	author={Abdellaoui, Mohammed and Baillon, Aur{\'e}lien and Placido, Laetitia and Wakker, Peter},
	journal={American Economic Review},
	volume={101},
	number={2},
	pages={695--723},
	year={2011}
}

@article{bade2015randomization,
	title={Randomization devices and the elicitation of ambiguity-averse preferences},
	author={Bade, Sophie},
	journal={Journal of Economic Theory},
	volume={159},
	pages={221--235},
	year={2015},
	publisher={Elsevier}
}

@article{manski2010rounding,
	title={Rounding probabilistic expectations in surveys},
	author={Manski, Charles F and Molinari, Francesca},
	journal={Journal of Business \& Economic Statistics},
	volume={28},
	number={2},
	pages={219--231},
	year={2010},
	publisher={Taylor \& Francis}
}

@book{Ok2007,
	author = {Ok, Efe A},
	title = {Real Analysis with Economic Applications},
	publisher = {Princeton University Press},
	address = {Princeton}, 
	year = {2007}
}

@article{georgalos2023higher,
	title={Higher order risk attitudes: new model insights and heterogeneity of preferences},
	author={Georgalos, Konstantinos and Paya, Ivan and Peel, David},
	journal={Experimental Economics},
	volume={26},
	number={1},
	pages={145--192},
	year={2023},
	publisher={Springer}
}

@article{bruhin2010risk,
	title={Risk and rationality: Uncovering heterogeneity in probability distortion},
	author={Bruhin, Adrian and Fehr-Duda, Helga and Epper, Thomas},
	journal={Econometrica},
	volume={78},
	number={4},
	pages={1375--1412},
	year={2010},
	publisher={Wiley Online Library}
}

@article{gul2015hurwicz,
	title={Hurwicz expected utility and subjective sources},
	author={Gul, Faruk and Pesendorfer, Wolfgang},
	journal={Journal of Economic Theory},
	volume={159},
	pages={465--488},
	year={2015},
	publisher={Elsevier}
}

@article{bleichrodt2023testing,
	title={Testing Hurwicz expected utility},
	author={Bleichrodt, Han and Grant, Simon and Yang, Jingni},
	journal={Econometrica},
	volume={91},
	number={4},
	pages={1393--1416},
	year={2023},
	publisher={Wiley Online Library}
}

@article{gaudecker2011heterogeneity,
	Author = {von Gaudecker, Hans-Martin and van Soest, Arthur and Wengstrom, Erik},
	Title = {Heterogeneity in risky choice behavior in a broad population},
	Journal = {American Economic Review},
	Volume = {101},
	Number = {2},
	Year = {2011},
	Pages = {664-94}
}

@article{quiggin1981risk,
	author = {Quiggin, John},
	title = {Risk perception and the analysis of risk attitudes},
	journal = {Australian Journal of Agricultural Economics},
	volume = {25},
	number = {2},
	pages = {160-169},
	doi = {https://doi.org/10.1111/j.1467-8489.1981.tb00393.x},
	year = {1981}
}

@article{gao2023behavioral,
	title={Behavioral welfare economics and risk preferences: A Bayesian approach},
	author={Gao, Xiaoxue Sherry and Harrison, Glenn W and Tchernis, Rusty},
	journal={Experimental Economics},
	volume={26},
	number={2},
	pages={273--303},
	year={2023},
	publisher={Springer}
}

@book{train2009discrete,
	title={Discrete choice methods with simulation},
	author={Train, Kenneth E},
	year={2009},
	publisher={Cambridge university press}
}

@InCollection{mcfadden1974conditional,
	Title                    = {Conditional Logit Analysis of Qualitative Choice Behaviour},
	Author                   = {D. McFadden},
	Booktitle                = {Frontiers in Econometrics},
	Publisher                = {Academic Press New York},
	Year                     = {1973},
	
	Address                  = {New York, NY, USA},
	Editor                   = {P. Zarembka},
	Pages                    = {105-142}
}

@article{agranov2022revealed,
	Author = {Agranov, Marina and Ortoleva, Pietro},
	Title = {Revealed Preferences for Randomization: An Overview},
	Journal = {AEA Papers and Proceedings},
	Volume = {112},
	Year = {2022},
	Pages = {426-30}
}

@article{armantier2016rich,
	title={The rich domain of risk},
	author={Armantier, Olivier and Treich, Nicolas},
	journal={Management Science},
	volume={62},
	number={7},
	pages={1954--1969},
	year={2016},
	publisher={INFORMS}
}

@article{brier1950verification,
	title={Verification of forecasts expressed in terms of probability},
	author={Brier, Glenn W},
	journal={Monthly Weather Review},
	volume={78},
	number={1},
	pages={1--3},
	year={1950}
}

@book{savage1972foundations,
	title={The Foundations of Statistics},
	author={Savage, Leonard J},
	year={1954},
	publisher={Wiley},
	address={New York}
}

@article{karni2009mechanism,
	title={A mechanism for eliciting probabilities},
	author={Karni, Edi},
	journal={Econometrica},
	volume={77},
	number={2},
	pages={603--606},
	year={2009},
	publisher={Wiley Online Library}
}

@article{baillon2018effect,
	title={The effect of learning on ambiguity attitudes},
	author={Baillon, Aur{\'e}lien and Bleichrodt, Han and Keskin, Umut and l’Haridon, Olivier and Li, Chen},
	journal={Management Science},
	volume={64},
	number={5},
	pages={2181--2198},
	year={2018},
	publisher={INFORMS}
}

@article{bianchi2019ambiguity,
	title={Ambiguity preferences and portfolio choices: Evidence from the field},
	author={Bianchi, Milo and Tallon, Jean-Marc},
	journal={Management Science},
	volume={65},
	number={4},
	pages={1486--1501},
	year={2019},
	publisher={INFORMS}
}

@article{muthukrishnan2009ambiguity,
	title={Ambiguity aversion and the preference for established brands},
	author={Muthukrishnan, AV and Wathieu, Luc and Xu, Alison Jing},
	journal={Management Science},
	volume={55},
	number={12},
	pages={1933--1941},
	year={2009},
	publisher={INFORMS}
}

@article{chateauneuf2007choice,
	title={Choice under uncertainty with the best and worst in mind: Neo-additive capacities},
	author={Chateauneuf, Alain and Eichberger, J{\"u}rgen and Grant, Simon},
	journal={Journal of Economic Theory},
	volume={137},
	number={1},
	pages={538--567},
	year={2007},
	publisher={Elsevier}
}

@article{manski2004measuring,
	title={Measuring expectations},
	author={Manski, Charles F},
	journal={Econometrica},
	volume={72},
	number={5},
	pages={1329--1376},
	year={2004},
	publisher={Wiley Online Library}
}

@article{gonzalez1999shape,
	title={On the shape of the probability weighting function},
	author={Gonzalez, Richard and Wu, George},
	journal={Cognitive Psychology},
	volume={38},
	number={1},
	pages={129--166},
	year={1999},
	publisher={Elsevier}
}

@article{haridon2019all,
	title={All over the map: A worldwide comparison of risk preferences},
	author={L’Haridon, Olivier and Vieider, Ferdinand M},
	journal={Quantitative Economics},
	volume={10},
	number={1},
	pages={185--215},
	year={2019},
	publisher={Wiley Online Library}
}

@article{Gneiting,
	author="T. Gneiting",
   TITLE = {Making and evaluating point forecasts},
 JOURNAL = {Journal of the American Statistical Association},
    VOLUME = {106},
      YEAR = {2011},
    NUMBER = {494},
     PAGES = {746--762},}

@article{haridon2018off,
	title={Off the charts: {Massive} unexplained heterogeneity in a global study of ambiguity attitudes},
	author={L’Haridon, Olivier and Vieider, Ferdinand M and Aycinena, Diego and Bandur, Agustinus and Belianin, Alexis and Cingl, Lubomir and Kothiyal, Amit and Martinsson, Peter},
	journal={Review of Economics and Statistics},
	volume={100},
	number={4},
	pages={664--677},
	year={2018},
	publisher={MIT Press One Rogers Street, Cambridge, MA 02142-1209, USA journals-info~…}
}

@article {stock,
author = {Stock, James H. and Wright, Jonathan H.},
title = {{GMM with weak identification}},
journal = {Econometrica},
volume = {68},
number = {5},
publisher = {Blackwell Publishers Ltd},
issn = {1468-0262},
url = {http://dx.doi.org/10.1111/1468-0262.00151},
doi = {10.1111/1468-0262.00151},
pages = {1055--1096},
year = {2000},
}

@article{hossain2013binarized,
	title={The binarized scoring rule},
	author={Hossain, Tanjim and Okui, Ryo},
	journal={The Review of Economic Studies},
	volume={80},
	number={3},
	pages={984--1001},
	year={2013},
	publisher={Oxford University Press}
}
}
\end{document}